\documentclass{aa}  

\usepackage{natbib}
\usepackage{graphicx}
\usepackage{txfonts}
\usepackage[colorlinks, citecolor=blue, linkcolor=blue]{hyperref}
\usepackage{color}
\usepackage{longtable}

\DeclareMathAlphabet{\pazocal}{OMS}{zplm}{m}{n}
\newcommand{\PriJ}{\pazocal{J}}
\newcommand{\PriU}{\pazocal{U}}

\newcommand{\serval}{{\tt serval}}
\newcommand{\tess}{{\it TESS}}

\makeatletter
\renewcommand*\aa@pageof{, page \thepage{} of \pageref*{LastPage}} 
\makeatother

\begin{document}

\title{The CARMENES search for exoplanets around M dwarfs}
\subtitle{Measuring precise radial velocities in the near infrared:\\
the example of the super-Earth CD\,Cet\,b%
\thanks{Based on observations collected at the Centro Astronómico Hispano Alemán (CAHA) at Calar Alto, Almer\'ia, Spain, operated jointly by the Junta de Andaluc\'ia and the Instituto de Astrof\'isica de Andaluc\'ia (CSIC).}\textsuperscript{,}%
\thanks{Based on observations collected at the European Southern Observatory, Paranal, Chile, under program 0103.C-0152(A), and La Silla, Chile, under programs 072.C-0488(E) and 183.C-0437(A).}}

\author{
    F.\,F.~Bauer \inst{1}
    \and M.~Zechmeister \inst{2}
    \and A.~Kaminski \inst{3}
    \and C.~Rodr\'{\i}guez L\'{o}pez \inst{1}
    \and J.~A.~Caballero \inst{4}
    \and M.~Azzaro \inst{5}
    \and O.~Stahl \inst{3}
    \and D.~Kossakowski \inst{6}
    \and A.~Quirrenbach \inst{3}
    \and S.~Becerril Jarque \inst{1}
    \and E.~Rodr\'{\i}guez \inst{1}
    \and P.\,J.~Amado \inst{1}
    \and W.~Seifert \inst{3}
    \and A.~Reiners \inst{2}
    \and S.~Sch{\"a}fer \inst{2}
    \and I.~Ribas \inst{7,8}
    \and V.\,J.\,S.~Béjar \inst{9,10}
    \and M.~Cort\'{e}s-Contreras \inst{4}
    \and S.~Dreizler \inst{2}
    \and A.~Hatzes \inst{11}
    \and T.~Henning \inst{6}
    \and S.\,V.~Jeffers \inst{2}
    \and M.~K\"urster \inst{6}
    \and M.~Lafarga \inst{7,8}
    \and D.~Montes \inst{12}
    \and J.\,C.~Morales \inst{7,8}
    \and J.\,H.\,M.\,M.~Schmitt \inst{13}
    \and A.~Schweitzer \inst{13}
    \and E.~Solano \inst{4}
}

\institute{
     Instituto de Astrof\'{\i}sica de Andaluc\'{\i}a (IAA-CSIC), Glorieta de la Astronom\'ia s/n, 18008 Granada, Spain\\
     \email{fbauer@iaa.es}
     \and Institut f\"{u}r Astrophysik, Georg-August-Universit\"{a}t, Friedrich-Hund-Platz 1, 37077 G\"{o}ttingen, Germany\\
     \email{zechmeister@astro.physik.uni-goettingen.de}
     \and Landessternwarte, Zentrum f\"{u}r Astronomie der Universt\"{a}t Heidelberg, K\"{o}nigstuhl 12, 69117 Heidelberg, Germany
     \and Centro de Astrobiolog\'{\i}a (CSIC-INTA), ESAC, camino bajo del castillo s/n, 28049 Villanueva de la Cañada, Madrid, Spain
     \and Centro Astron\'{o}mico Hispano-Alem\'{a}n (CSIC-Junta de Andaluc\'{\i}a), Observatorio Astron\'{o}mico de Calar Alto, Sierra de los Filabres, 04550 G\'{e}rgal, Almer\'{\i}a, Spain
     \and Max-Planck-Institut f\"{u}r Astronomie, K\"{o}nigstuhl 17, 69117 Heidelberg, Germany
     \and Institut de Ci\'{e}ncies de l’Espai (ICE, CSIC), Campus UAB, c/de Can Magrans s/n, 08193 Bellaterra, Barcelona, Spain
     \and Institut d’Estudis Espacials de Catalunya (IEEC), C/Gran Capità 2-4, 08034 Barcelona, Spain
     \and Instituto de Astrof\'{\i}sica de Canarias, V\'{\i}a L\'{a}ctea s/n, 38205 La Laguna, Tenerife, Spain
     \and Departamento de Astrof\'{\i}sica, Universidad de La Laguna, 38026 La Laguna, Tenerife, Spain
     \and Th\"{u}ringer Landessternwarte Tautenburg, Sternwarte 5, 07778 Tautenburg, Germany
     \and Facultad de Ciencias F\'{\i}sicas, Departamento de F\'{\i}sica de la Tierra y Astrof\'{\i}sica \& IPARCOS-UCM (Instituto de F\'{\i}sica de Part\'{\i}culas y del Cosmos de la UCM), Universidad Complutense de Madrid, 28040 Madrid, Spain
     \and Hamburger Sternwarte, Gojenbergsweg 112, 21029 Hamburg, Germany
}

\date{Received 26 March 2020 / Accepted 26 May 2020}

\abstract{The high-resolution, dual channel, visible and near-infrared spectrograph CARMENES offers exciting opportunities for stellar and exoplanetary research on M dwarfs. In this work we address the challenge of reaching the highest radial velocity precision possible with a complex, actively cooled, cryogenic instrument, such as the near-infrared channel. We describe the performance of the instrument and the work flow used to derive precise Doppler measurements from the spectra. The capability of both CARMENES channels to detect small exoplanets is demonstrated with the example of the nearby M5.0\,V star CD\,Cet (GJ 1057), around which we announce a super-Earth ($4.0\pm0.4\,M_\oplus$) companion on a 2.29\,d orbit. 
}


\maketitle

\section{Introduction}

Our ability to detect exoplanets with radial velocities (RVs) has constantly grown with advances in technology and methodology. Hard-won early planetary discoveries, such as 51\,Peg\,b \citep{51Peg}, have now become easily detectable, and the field steadily approaches the domain of small, rocky planets (e.g., \citealp{Proxima}; \citealp{Teegarden}; \citealp{GJ1061}). The detection of Earth analogs orbiting in the habitable zone may currently be beyond our reach in G stars for a number of reasons. However, we can probe this interesting region of the exoplanet space with M dwarfs. 
Most of the closest neighbors of the Sun are M dwarfs, which are the most numerous stars in the Galaxy \citep{2006AJ....132.2360H}. M dwarfs seem to harbor at least one or two planets per star \citep{Ga16,HU19}. Therefore, many, if not most, planets in the Galaxy orbit M dwarfs. If we want to understand planet formation processes, planet compositions, and habitability, M dwarfs in our vicinity are a good starting point. 
Moreover, small, rocky planets orbiting within the habitable zone of M stars are indeed already within our detection ability. Although habitability is still vaguely defined \citep{HZ_Kasting,Ta07}, if life develops under a broad range of conditions, inhabited worlds could be numerous, close by, and already detectable for us.

As a result, M dwarfs have become the focus of many exoplanetary studies. As M dwarfs are intrinsically faint and their spectral energy distribution peaks in the near infrared, instrumentation had to adapt over the last years. This drove the extension from blue-sensitive spectrographs such as HARPS \citep{Mayor2003Msngr.114...20M} to near-infrared instruments, such as CARMENES\footnote{Calar Alto high-Resolution search for M dwarfs with Exoearths with Near-infrared and optical \'Echelle Spectrographs.} \citep{CARMENES_instrument_overview,CARMENES_near_infra_read_spectra}, GIANO \citep{GIANO}, iSHELL \citep{iSHELL} the successor of CSHELL \citep{CSHELL}\footnote{Cryogenic Echelle Spectrograph}, SPIRou\footnote{SPectropolarim\`etre InfraROUge.} \citep{Spirou}, IRD\footnote{InfraRed Doppler for the Subaru telescope.} \citep{IRD14}, and HPF\footnote{Habitable zone Planet Finder.} \citep{HPF}, which are already in operation or those that will be operative soon, such as NIRPS\footnote{Near Infra Red Planet Searcher} \citep{NIRPS17}, CRIRES+\footnote{CRyogenic high-resolution InfraRed Echelle Spectrograph Plus.} \citep{CRIRES_plus}, or Veloce Rosso \citep{Veloce_Rosso}. Making this move, however, requires solving several technical issues.
Near-infrared spectrographs need to be actively thermally stabilized to a high degree of precision \citep{Carmenes_tank, HPF_tank, Spirou_tank}.
Furthermore, HgCdTe hybrid near-infrared detectors work differently from optical charge-coupled devices (CCDs) \citep{Bec19}.
Last but not least, the calibration procedure requires sources and methods that are adapted to the new wavelength regime \citep{NIR_FPs_Schaefer, FP_Bauer, Luis, FP_Schaefer, CARMENES_near_infra_read_spectra}.

CARMENES was among the first operational near-infrared instruments to have taken on all these challenges. The system has finished its fourth year in operation and has proven its value for exoplanetary science through new detections and confirmations of planets, and through investigations of their atmospheres \citep[e.g.,][]{Sa18, No18, Barnard_c, Teegarden, GJ3512}. 


In this paper, we present our experience and provide insight for scientists outside the consortium working with CARMENES data. To showcase the capabilities of CARMENES, we also announce the discovery of a super-Earth orbiting in the temperate zone of an M5.0\,V star.
Section~\ref{sec:CARMENES} summarizes the main characteristics of both spectrograph channels and presents their stability limits, along with insights about their causes. In Sect.~\ref{sec:RV_computation} we show our recipe to derive precise RVs with the near-infrared channel of CARMENES, and how they compare to those of the visible. In Sect.~\ref{sec:star} we apply our RV methods and present the discovery of the temperate super-Earth CD\,Cet\,b at visible and near-infrared wavelengths. We summarize and discuss the abilities of CARMENES in Sect.~\ref{sec:discussion}.


\section{CARMENES in operation}\label{sec:CARMENES}

\subsection{CARMENES in a nutshell}
\label{sec:CARMENES_summary}

CARMENES \citep{CARMENES_instrument_overview,CARMENES_near_infra_read_spectra} consists of two channels: the optical channel (from now on VIS), which covers the range 5200--9600\,\AA{} at a resolution of $R=94\,600$, and the near-infrared channel (from now on NIR), which covers 9600--17\,100\,\AA{} at $R=80\,400$. 
Both spectrograph channels are located at the 3.5\,m Calar Alto telescope at the Centro Astronómico Hispano-Alem\'an (CAHA). They are fed from the Cassegrain focus via fiber links containing sections with octagonal cross-sections, which improves the scrambling, that is, the stability of the fiber output when the input varies due to telescope tracking errors or variable seeing \citep{2014SPIE.9151E..52S}. The entrance apertures of the spectrographs are formed by images of the fibers, sliced into two halves to allow for the high resolving power \citep{2012SPIE.8446E..33S}. For maximum temperature stability, all optical elements are mounted on solid, aluminum optical benches and placed inside vacuum tanks, which in turn are located in three separate temperature-controlled rooms within the coud\'e room below the telescope (one for each spectrograph channel, and one for both calibration units).

A major difference between the two CARMENES spectrograph channels is their temperature stabilization concept. The VIS channel is passively stabilized. It thus follows the temperature of the ambient chamber, albeit slowly due to its thermal inertia and insulation.
In contrast, the NIR channel is actively stabilized \citep{Carmenes_tank}. Efficient observations beyond $1\,\mu$m require minimizing the thermal background radiation by cooling the instrument. Thus, the NIR channel is fed by $\mathrm{N_2}$ gas at about 140\,K. This operating temperature is set by the requirement that the thermal background seen by the detector, which is sensitive to about 2.4\,$\mu$m, is lower than the dark current. Keeping instrumental changes small is crucial, as milli-Kelvin temperature fluctuations translate into absolute drifts on the order of a few m\,s$^{-1}$. Maintaining the instrumental temperature ultra-stable by a cooling circuit is, thus, among the main challenges of obtaining high-precision RVs in the near-infrared. Hence, substantial engineering efforts during the first year of operations (2016) with CARMENES were dedicated to optimizing the NIR cooling system (Sect.~\ref{sec:NIR_performance}).

Wavelength calibration is another factor critical for obtaining high-precision RVs even with stabilized spectrographs. Hence, the VIS calibration unit is equipped with three types of hollow-cathode lamps (thorium-neon, uranium-neon, and uranium-argon). For the NIR channel we use only an uranium-neon lamp, because thorium lamps deliver too few lines \citep{2014ApJS..211....4R} and argon emits strong features between 0.96\,$\mu$m and 1.71\,$\mu$m. In addition, both spectrograph channels have their own dedicated Fabry-P\'erot (FP) etalons, which are used for simultaneous drift measurements during the night \citep{FP_Schaefer}, and also for the construction of the daily wavelength solution \citep{FP_Bauer}.
Below, we examined the whole data set built with all calibration observations carried out during the past four years to present an overview of the VIS and NIR channel performance.

\subsection{VIS channel performance}
\label{sec:VIS_performance}

\begin{figure} 
     \includegraphics[width=\linewidth]{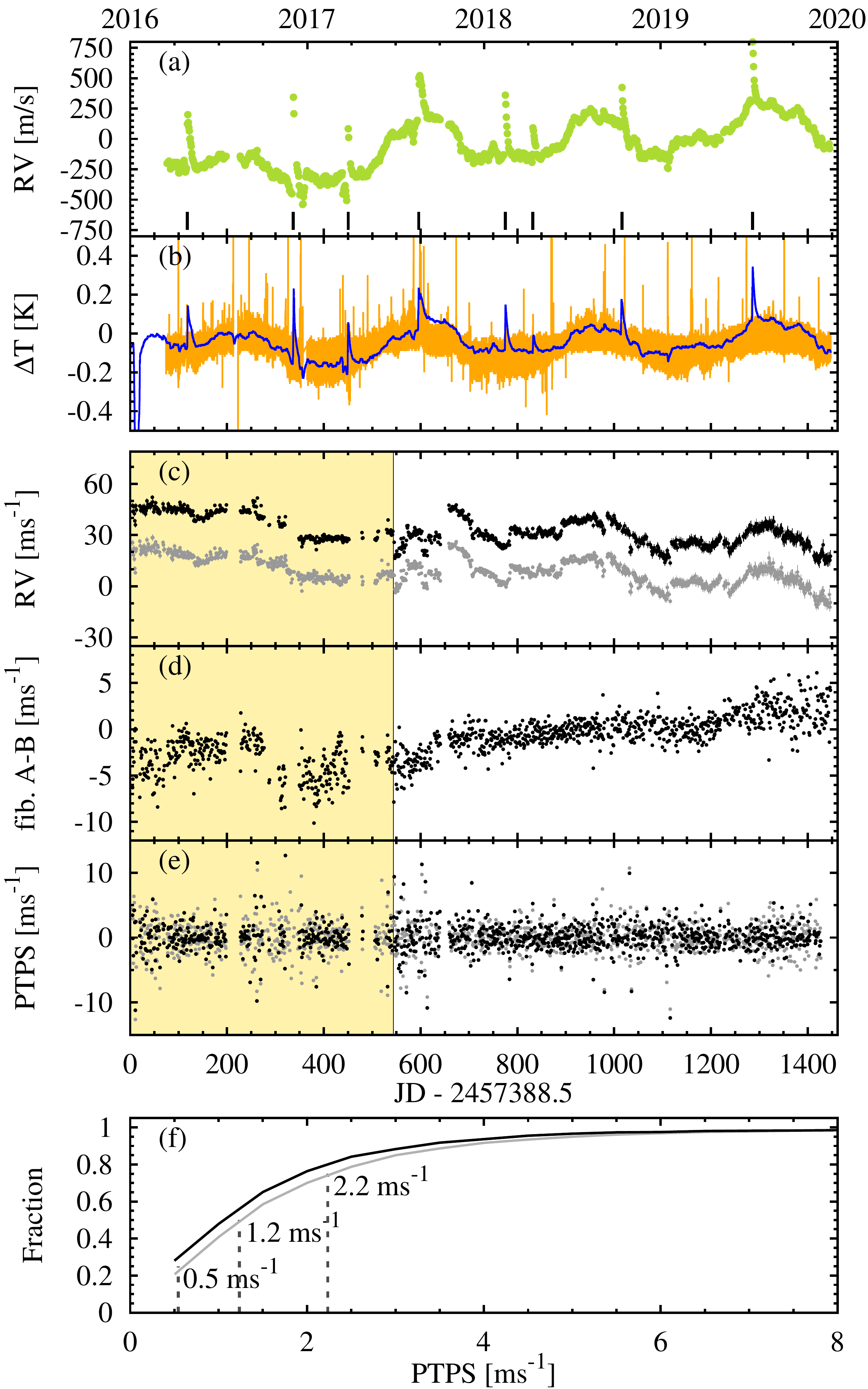} 
     \caption{{\it Panel~($a$):} absolute RV drift of CARMENES VIS over four years. Detector cryostat sorption pump re-generations are marked with black ticks. 
     {\it Panel~($b$):} temperature of the CARMENES VIS room (orange) and of the VIS optical bench (blue).
     {\it Panel~($c$):} Time series of the VIS FP. Object fiber~A (black) and reference fiber~B (gray). 
     {\it Panel~($d$):} relative drift between fiber~A and fiber~B. 
     {\it Panel~($e$):} PTPS computed from panel ($c$). Shaded area: data taken prior to the FP coupling change. 
     {\it Panel~($f$):} cumulative histogram of PTPS marked with the 25\,\%, 50\,\%, 75\,\% quantiles (vertical dashed lines). The median precision of the VIS channel is $1.2$~m\,s$^{-1}$.
     }
     \label{fig:joint_VIS} 
\end{figure}



Except for a few days when large maintenance programs are carried out, CARMENES is calibrated every afternoon a few hours before first observations take place at Calar Alto. The wavelength solution is recomputed for each individual calibration set, that is, daily. To measure the absolute instrument drift, we computed the median difference between each new CARMENES wavelength solution and the first one obtained at the beginning of the survey. We plot this absolute instrumental drift together with the VIS room temperature and the VIS optical bench temperature in Fig.~\ref{fig:joint_VIS} (panels $a$ and $b$).

The main source of long-term instrumental drifts in the VIS channel are seasonal temperature variations inside the spectrograph room. These variations ($\pm 0.2$\,K) are well within the precision that the climatic control inside the room is able to achieve. These changes in room temperature then propagate toward the optical bench. Slight alignment changes due to thermal expansion cause seasonal instrumental drifts of about 400--600\,m\,s$^{-1}$ peak to peak. The translation from optical bench temperature change to instrumental drift is about 2\,\,m\,s$^{-1}$\,mK$^{-1}$.


Larger excursions (indicated by black ticks in
Fig.~\ref{fig:joint_VIS}, panel $a$) 
are related to infrequent vacuum re-generations inside the VIS channel detector cryostat. In general, the instrument quickly settles after such a maintenance action.


To test the performance of our wavelength calibration strategy, we investigated the RV drift of our FP spectra.
While the FP spectra assists our wavelength solution during the calibration process, each individual wavelength solution is anchored to the standard hollow-cathode lamps. This means that slow intrinsic Doppler-like drifts in the FP spectra are calibrated out from night to night, but these drifts can be computed by treating the FP spectra like stellar spectra. For this exercise, we selected one FP spectrum taken during each daily calibration run and derived RVs as we do for the stars. We used {\serval} \citep{Zechmeister2018A&A...609A..12Z} to perform a least-squares matching between an observed spectrum and a high signal-to-noise ratio (S/N) template.
In Fig.~\ref{fig:joint_VIS} (panel $c$), we present the RV curve for the VIS FP. The shaded area indicates FP spectra taken prior to a change in the coupling of the FPs into the calibration units, which resulted in a more homogeneous illumination of the calibration fibers, but led to a jump in the FP drift curve of about 100.5\,m\,s$^{-1}$, for which we corrected in this plot.

The FP RVs show a slowly varying RV drift with a pattern of seasonal variations much smaller than the absolute spectrograph drift in panel ($a$) of Fig.\,\ref{fig:joint_VIS}.  These variations are probably related to slow changes in the alignment of the FP system (see \citealp{NIR_FPs_Schaefer} and \citealp{FP_Schaefer} for more details). The long term relative drift between both fibers is re-calibrated every night by a new wavelength solution. It can, however, be made visibly by subtracting the FP RVs measured in both fibers (panel~$c$) from each other as shown in panel ($d$). We do find a small long-term differential drift, but the seasonal temperature modulations observed in the absolute drift (top panel $a$ of Fig.~\ref{fig:joint_VIS}) are similar for both fibers. Thus, there is no significant seasonal relative drift between the fibers.  The day-to-day jitter around the smooth FP drift of panel ($c$) represents the precision of our VIS channel, including the calibration strategy. We quantified this mostly day-to-day scatter by computing the point-to-point scatter (PTPS, see panel $e$ of Fig.~\ref{fig:joint_VIS}) and the cumulative histogram of the PTPS (panel $f$ of Fig.~\ref{fig:joint_VIS}). We conclude that the median RV precision reached with the current calibration and data reduction procedures of CARMENES VIS is $1.2$~m\,s$^{-1}$.

\subsection{NIR channel performance}\label{sec:NIR_performance}

\begin{figure} 
     \includegraphics[width=\linewidth]{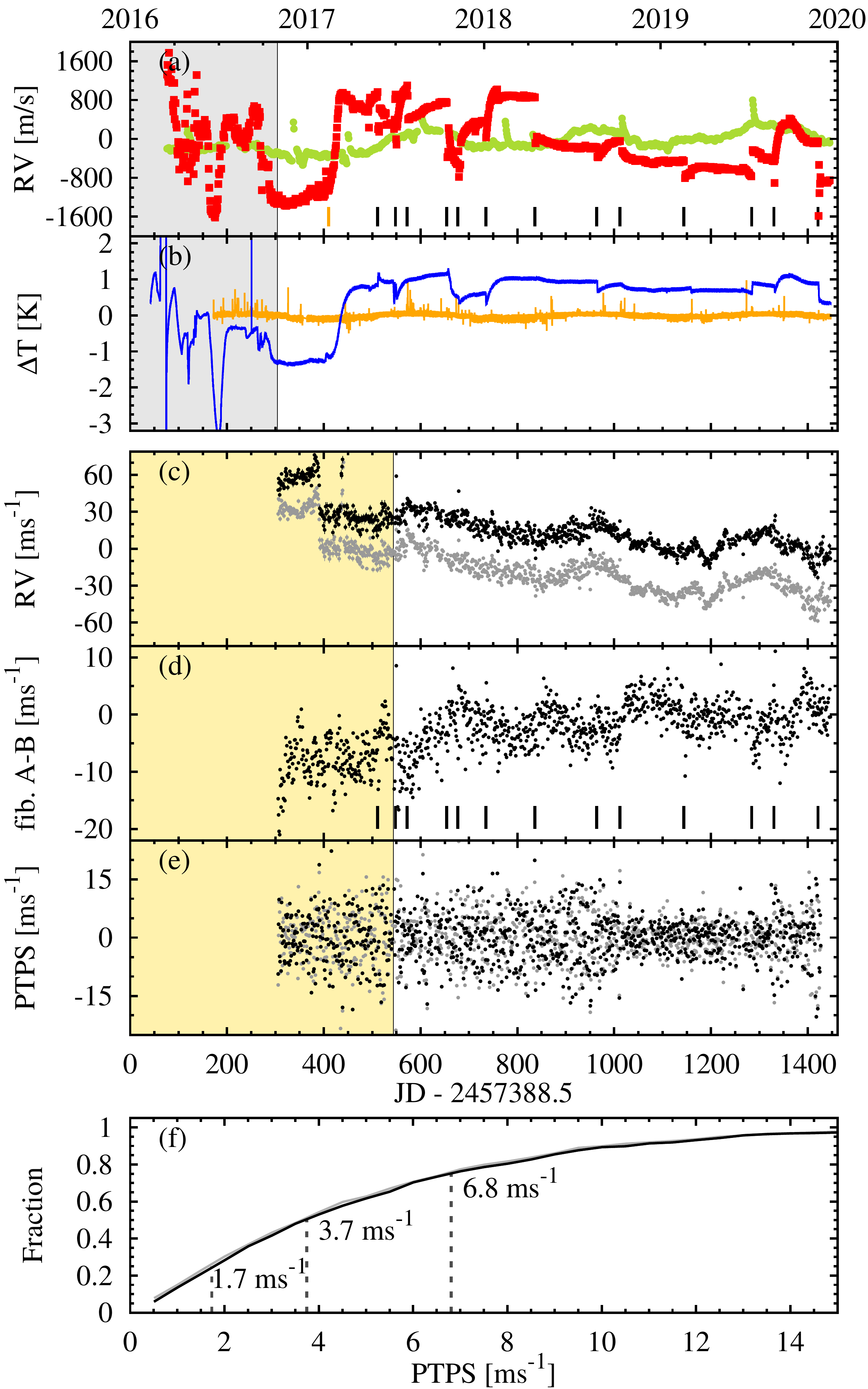} 
     \caption{Same as Fig.~\ref{fig:joint_VIS}, but for CARMENES NIR. For comparison the VIS instrumental drift is over-plotted in green. The orange tick indicates a planned temperature rise of the NIR channel, while the black ticks mark larger maintenance operations. 
     The gray shaded area indicates an initial phase of poor thermal stability. The median precision of the NIR channel is $3.7$~m\,s$^{-1}$.
     }
     \label{fig:joint_NIR} 
\end{figure}



We begin the instrument performance analysis of CARMENES NIR by elucidating the instrumental drift with the aid of Fig.~\ref{fig:joint_NIR}.  As a result of an initial phase of poor thermal stability (gray shaded area in top panels $a$ and $b$ of Fig.~\ref{fig:joint_NIR}), the nominal operations of CARMENES NIR began on 2016-11-01. We find the NIR channel response to optical bench temperature changes to be similar to that of the VIS channel (about 2\,\,m\,s$^{-1}$\,mK$^{-1}$), while its peak-to-peak instrumental drift of about 3200\,m\,s$^{-1}$ is substantially larger than that of the VIS channel.
Most instrumental changes in the NIR channel occur, however, during maintenance operations of the complex active cryogenic cooling system (marked by black ticks in 
panel $a$ of Fig.~\ref{fig:joint_NIR}). 
During nominal operations the instrumental drift of the VIS and NIR channels are comparable. 
We repeated the FP analysis of Sect.~\ref{sec:VIS_performance} for NIR. The resulting RV curve, differential drift, PTPS, and cumulative histogram are plotted in panels ($c$) to ($f$) of Fig.~\ref{fig:joint_NIR}. 
When comparing the FP RV curves between the VIS and NIR channels
(panels $c$ of Figs.~\ref{fig:joint_VIS} and~\ref{fig:joint_NIR}, respectively), 
one may notice the similarities in the trends. The reason for both FP systems drifting in a rather similar way is likely that they share the same temperature control unit. Thus, they are not completely independent, and any change in coolant temperature affects the VIS and NIR FPs alike.  In the relative drift, shown in panel ($d$) of Fig.~\ref{fig:joint_NIR}, we do find indications for a link between absolute spectrograph and relative fiber drift in the NIR channel. Cooling system maintenance actions (marked by black ticks -- and one orange tick) are typically the onset of differential fiber changes. The precision of the NIR channel, plotted in panels ($e$) and ($f$) in Fig.~\ref{fig:joint_NIR}, is about three times worse than that of the VIS channel. Over the 2017--2019 interval we obtain a median precision of $3.7$~m\,s$^{-1}$ for CARMENES NIR.
However, having gained experience with the stabilization of a cryogenic instrument, we find that there is still room for improvements in CARMENES NIR. As seen around JD~2\,458\,400 in Fig.~\ref{fig:joint_NIR}, panel (e), the precision of the NIR instrument improved from a median of $4.48$\,m\,s$^{-1}$ to $2.36$\,m\,s$^{-1}$. The reason was a maintenance action improving the insulation of all pipes feeding the instrument with N$_2$ cooling gas. Thus, the coolant arriving in the instrument experiences less warm up and the feedback loop to thermally stabilize the instrument works more accurately.

\section{Radial velocity computation}\label{sec:RV_computation}

\begin{figure*} 
     \includegraphics[width=\linewidth]{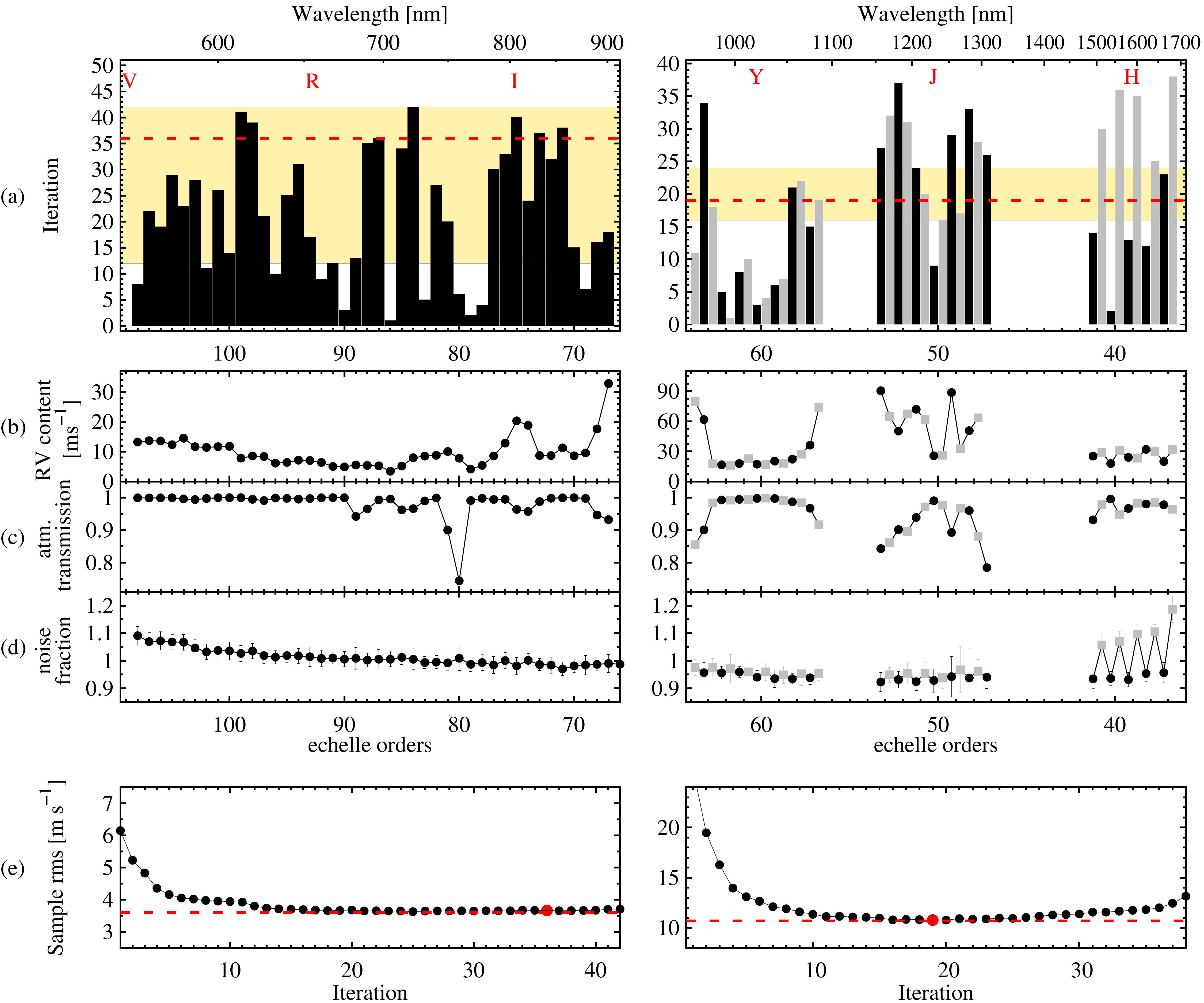} 
     \caption{{\it Panel ($a$):} order selection procedure to minimize the sample rms of the VIS ({\em left}) and NIR ({\em right}) channels. Each bar represents the number of iterations that the algorithm takes until this order is used in the RV determination. Half orders that fall on the blue detector of the NIR channel are marked in black, while red half orders are colored gray. The iteration limit at which the minimum sample rms is reached is indicated by the red dashed horizontal line. The light orange shaded area marks orders which, when added or rejected, keep the sample rms within 0.2\,m\,s$^{-1}$.
     {\it Panel ($b$):} RV content per spectral (half) order.
     {\it Panel ($c$):} average atmospheric transmission per spectral (half) order.
     {\it Panel ($d$):} ratio between measured noise in the spectra and pixel uncertainty (photon plus read out).
     {\it Panel ($e$):} median sample rms as a function of the number of selected orders for the VIS ({\em left}) and NIR ({\em right}) channels. 
    The median sample rms attains its minimum, indicated by the red dashed horizontal line, at the iteration marked by the big red dot (VIS: 35 iterations at rms 3.6\,m\,s$^{-1}$, NIR: 19 iterations at rms 10.7\,m\,s$^{-1}$).}
     \label{fig:orderselection}
\end{figure*}


Reaching an RV precision at the m\,s$^{-1}$ level requires the use of thousands of spectral lines in order to collect sufficient information \citep{Bo01}. Hence, all state-of-the-art RV instruments employ the cross-dispersed \'{e}chelle design \citep{ELODIE}, which re-formats a large wavelength range at high resolution onto a square or rectangular detector. RVs are typically measured order-by-order and are then combined into a final measurement value \citep{Zechmeister2018A&A...609A..12Z}.
Particularly in the near-infrared range, however, spectral information is unevenly distributed over spectral orders caused by differences in the distribution of flux, stellar molecular absorption, and telluric (water) absorption, not counting physical gaps between units in the detector array (see \citealp{Reiners2018_324_stars} and \citealp{Reiners2020}). We expect spectral regions with low information content to yield RVs that are more strongly affected by systematics, such as weak and unmasked telluric lines or detector defects. In this section, we aim to optimize the CARMENES RV precision by identifying those spectral orders that provide the most precise RV data and those that exhibit small, unseen systematics.

We ran {\serval} \citep{Zechmeister2018A&A...609A..12Z} on all M stars in the CARMENES survey and started with an investigation of the orders. The NIR channel contains a mosaic of two 2k$\times$2k detectors arranged side by side with a gap, whose width corresponds to about $140$ pixels. RVs are, therefore, measured separately for the two detectors, that is, for half-orders, by {\serval}. This provides 38 wavelength chunks (excluding those within the telluric bands), each delivering an independent RV measurement ({\serval} rvo files). In contrast, the VIS channel contains 61 individual orders on a single monolithic 4k$\times$4k CCD. The six reddest orders are cut off by the dichroic beam splitter in the front end, which separates the light for the VIS and the NIR channels, and they are thus not used to derive RVs. We also rejected the last four orders before the beam splitter cutoff because of strong telluric bands and the ten bluest orders, because they are cropped and typically yield low signal-to-noise ratios (S/N) for M stars. Hence, the VIS channel provides 41 independent RV measurements.

The task at hand is now to identify which of these orders are the most useful to maximize the RV precision, and which ones systematically bias the final RV data. The systematics contributed by single orders is likely below the m\,s$^{-1}$ level. Thus, regions strongly affected by unmasked tellurics or other uncorrected effects can hardly be identified in the data of a single star. However, we expect the spectral orders contributing systematics to be the same in a large sample of similar stars. We used the following iterative approach making use of the large CARMENES target sample to identify the useful orders:

\begin{enumerate}
    \item For the individual wavelength chunks, we compute the root-mean-square (rms) scatter for each star in the entire CARMENES guaranteed time observations sample \citep{Reiners2018_324_stars}, after correction for nightly offsets using the nightly zero points \citep[NZP;][]{CARMENES_one_Trifon,HARPS_NZP,Lev_NZPs}. We rank the wavelength chunks by the median rms scatter derived from the entire CARMENES sample.
    \item We select the wavelength chunk with the lowest sample rms and remove it from the pool of available RV measurements.
    \item Each remaining RV measurement in the pool is combined with the (half-) orders that were already selected. We use inverse squared RV errors as weights.
    \item Again we draw the chunk from the pool that, in combination with those already drawn, results in the lowest sample rms.
    \item We continue at point three until no (half-) orders are left in the pool.
\end{enumerate}
It is important to note that we excluded stars with strong intrinsic signals (rms $>$ 100\,m\,s$^{-1}$) from the sample when examining the orders.

Panel ($a$) of Fig.~\ref{fig:orderselection} illustrates graphically the entire process for the VIS and NIR channels. Our final selection of useful (half-) orders contains the set that minimizes the median CARMENES sample rms (bottom panel of Fig.~\ref{fig:orderselection}). For the VIS channel almost all orders were selected (35 of 41). The remaining six orders that the procedure rejects, however, do not impact the RV precision significantly (0.05\,m\,s$^{-1}$) and are, therefore, kept when computing the RVs. For the NIR channel the situation is different, and we obtain the best precision with only 19 half-orders, which are used in what follows to derive homogeneously the RVs for all our sample stars.

We suspect rejected orders to contain systematics. Therefore, orders that are identified to impact RV performance significantly should pass a number of consistency tests. First, we tested the orders selected in Fig.~\ref{fig:orderselection} against 10\,000 random order combinations. None of these random draws outperforms the selection of our algorithm and, on average, random selection reaches a sample rms of 14.2\,m\,s$^{-1}$ in the NIR channel (10.7\,m\,s$^{-1}$ with the selection algorithm). Hence, the presented order selection algorithm identifies contaminated orders reliably. For the second test we split our sample by survey year. There should be no significant change in rejected orders over time and indeed orders flagged as bad (above the 0.2\,m\,s$^{-1}$ tolerance zone) are consistent overall. In a third test, we split our sample in two equally sized subsamples and ran the algorithm separately on both. Again we do not find significant differences.

The stable results likely indicate the presence of minuscule, hidden systematics in the orders that are rejected. In panels ($b$) to ($d$) in Fig.~\ref{fig:orderselection} we present the most likely candidates. Most of the NIR RV information in M~dwarfs is located in the $Y$ and $H$ bands, while there exists a paucity of spectral features in the $J$ band (see panel $b$, as well as \citealp{Reiners2018_324_stars} and \citealp{Reiners2020}). Telluric contamination is also less severe in the $Y$ and $H$ bands as compared to the $J$ band so that more pixels are available for RV measurement after masking atmospheric lines of Earth (panel $c$). Thus, we expect the $Y$ and $H$ bands to be most reliable for RV determination, while we expect the $J$ band to contribute more systematics, possibly as a result of unmasked tellurics. 

In addition to those astrophysical limitations, the red CARMENES NIR detector suffers from a hot corner that limits its performance in the $H$ band. In panel ($d$) we measured the pixel to pixel scatter in the spectra of a fast rotating F5\,V star and related it to expectations from photon and read out noise. The F5\,V star was chosen because it delivers relatively constant S/N values (slightly decreasing toward the blue end of the VIS channel) allowing for a fair rating of underlying detector noise. Values close to unity indicate that the noise in the spectra is explained well by photon and readout noise. Values above unity hint toward additional, unaccounted noise contributions. The difference between the red and blue detector performance toward our reddest spectral orders in CARMENES NIR is clearly visible, explaning why half of the $H$ band is rendered unusable (with the current data reduction) for high precision RV measurements.

Currently our selection algorithm works with relatively wide wavelength grids. We investigated the stellar spectra to understand if a smaller grid would safe parts of the rejected orders and further improve RV precision in CARMENES. In the VIS channel, no orders suffer strong contamination. In the NIR channel, there is very little stellar information to be saved in the $J$ band and increased detector noise on the red edge of the red CARMENES NIR detector is evenly distributed along the orders. Hence, we currently do not expect significantly improved RVs when using finer wavelength bins in both channels. In the future, however, when telluric contamination is modeled and removed from the spectra, a finer grid may help identifying telluric residuals and thus lead to precision improvements. Other instruments, especially those aiming at sub m\,s$^{-1}$ precision, may as well benefit from a reduction of systematic biases on a suborder scale.

CARMENES currently measures the drift globally and we tested the impact of using all, or only the set of selected good orders, to derive and correct the instrument drift in the NIR channel. Both drift measurements are consistent, as the median difference (0.3\,m\,s$^{-1}$) is comparable to the median error of the drift (0.42\,m\,s$^{-1}$). Tellurics are weaker in the calibration fiber and detector noise is less problematic for the FP because it is brighter than most of our stars. Thus, drift measurements are less vulnerable to the systematics that impact stars.



\begin{figure} 
    \includegraphics[width=\linewidth]{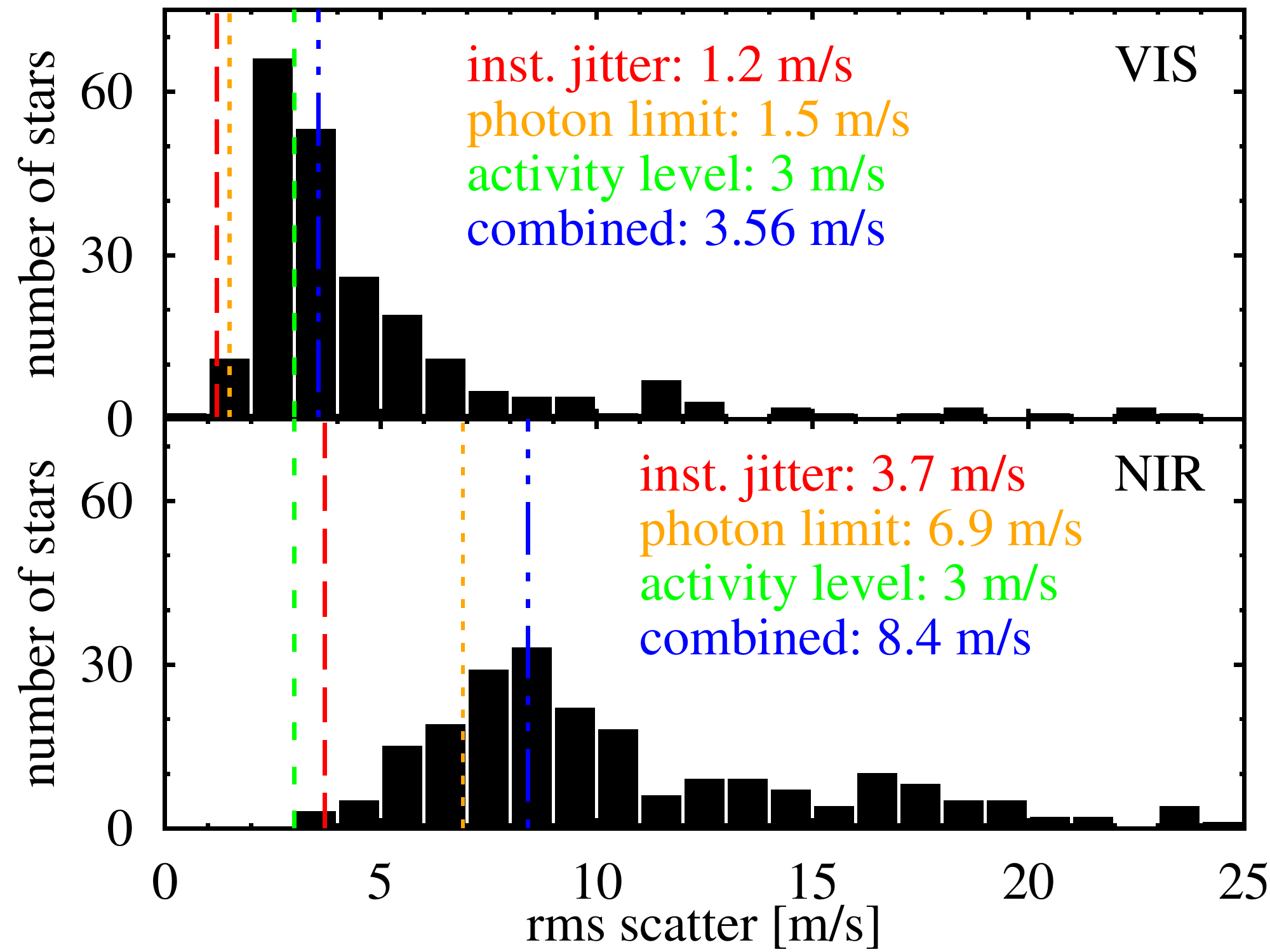} 
    \caption{Histogram of rms scatter exhibited by the CARMENES sample stars observed by the CARMENES VIS channel ({\it top}) and the NIR channel ({\it bottom}). Vertical lines mark the instrument jitter (red long-dashed), the median photon noise of the sample (orange dotted), the average M dwarf activity level (green short-dashed) and the quadratically summed jitter (blue dash-dotted). 
     }
    \label{fig:rv_stat}
\end{figure}

We now proceed to the evaluation of the CARMENES RV performance on sky. In Fig.~\ref{fig:rv_stat}, we show the rms scatter histogram for our M dwarf sample for the VIS and NIR channels. In line with previous RV surveys \citep[e.g.,][]{Wright2005PASP..117..657W,2009A&A...505..859Z,Bonfils2013A&A...549A.109B}, we find that the typical RV rms of our M dwarfs observed by the VIS channel is around 3--5\,m\,s$^{-1}$. This is significantly above our median photon noise limited RV precision of around 1.5\,m\,s$^{-1}$ (as computed by {\serval}). Hence, we attribute most of this noise floor to stellar activity jitter. 
The rms distribution in the NIR channel peaks at significantly higher values, with its maximum located around 7--9\,m\,s$^{-1}$. This distribution can be explained by the lower RV content in the parts of the NIR channel spectra that are useful for precise RV determination (median limit of $6.9$\,m\,s$^{-1}$), the higher instrumental jitter ($3.7$\,m\,s$^{-1}$, see Sect.~\ref{sec:NIR_performance}), and stellar activity. After the better N$_2$ pipe insulation improved the NIR channel stability, the measured stellar RVs improved in rms by about 1\,m\,s$^{-1}$. This is in line with expectations when combining all jitter terms. Thus, the relative importance of NIR RVs as compared to those of VIS has improved since September 2018. 

\section{An application to CD\texorpdfstring{\,}{~}Cet}\label{sec:star}

\subsection{Host star}

\object{CD\,Cet} (GJ~1057, Karmn J03133+047) is a bright \citep[$J \approx 8.8$\,mag,][]{2MASS06}, nearby \citep[$d \approx$ 8.6\,pc,][]{Gaia18b}, M5.0\,V star \citep{1984AJ.....89.1229R, 1994AJ....107..333K, 1996AJ....112.2799H,  2013AJ....145..102L, 2015ApJ...802L..10T} discovered in the Lowell proper motion survey of \citet{1961LowOB...5...61G}, who tabulate it as G~77--31. 
The mid-type M dwarf is relatively quiet, which is  reflected by a low chromospheric activity level as indicted by measurements of $\log R'_{\rm HK}$ between $-5.16$ \citep{Activity_catalogue} and $-5.52$ \citep{HARPS_activity}. This is in good agreement with its long rotational period of about $126.2$\,d determined from MEarth photometry \citep{Newton_rotation}, low rotational velocity \citep{Reiners2018_324_stars}, absent H$\alpha$ emission \citep[][see also Sect.~\ref{sec:Act_ana}]{2018A&A...614A..76J, Schoefer2019A&A...623A..44S}, and its non detection in volume-limited {\em ROSAT} surveys of extreme ultraviolet/soft X-ray emission from all stars within 10\,pc \citep{1994ApJS...93..287W, 2013MNRAS.431.2063S}.

CD\,Cet was the subject of high-resolution lucky imaging searches by \citet{2012ApJ...754...44J} and \citet{multiplicity_Miriam}, who impose stringent upper limits on the presence of stellar and substellar companions down to 8\,mag fainter than CD~Cet in the red optical at angular separations from 0.2 to 5.0\,arcsec. 
We also applied the same methodology of \citet{2015A&A...583A..85A} and searched for wider common proper motion companions within the {\em Gaia} DR2 completeness and up to 2\,deg, but without success\footnote{We verified our search with the recovery of \object{HD~20369}\,AB (A~2032), the only star pair tabulated by the Washington Double Star catalog \citep{2001AJ....122.3466M} in several degrees around CD~Cet.}.
To sum up, CD~Cet is a single star.

Table~\ref{tab:Parameters} summarizes some basic stellar parameters mostly compiled from the CARMENES M dwarf input catalog \citep[Carmencita,][]{Carmencita} and following addenda by \citet{Pass19} and \citet{Schw19}.
Our astrophysical stellar parameters ($T_{\rm eff}$, $\log{g}$, [Fe/H]), based on CARMENES VIS+NIR channel spectra, reasonably match previous determinations by \citet{2013AJ....145..102L}, \citet{2013A&A...556A..15R, 2018A&A...620A.180R},  \citet{2014MNRAS.443.2561G}, \citet{2014AJ....147...20N}, and  \citet{2015ApJS..220...16T}. 
We recomputed the galactocentric space velocities as in \citet{Cortes2016UCM-PhD} with the latest absolute radial velocity of \citet{Laf20}, which is identical within uncertainties to that of \citet{Reiners2018_324_stars} and supersedes previous determinations \citep{2014AJ....147...20N,  2015ApJS..220...16T, 2015ApJ...812....3W}.
With the tabulated $UVW$ values, CD~Cet belongs to the Galactic thin disk population, from which we estimate an age between 1\,Ga and 5\,Ga \citep{Cortes2016UCM-PhD}.

\begin{table}
\centering
\caption{\label{tab:Parameters} Stellar parameters of \object{CD\,Cet}.}
\begin{tabular}{@{}lcr@{}}
\hline\hline
\noalign{\smallskip}
Parameters & Value & Ref.\\
\noalign{\smallskip}
\hline
\noalign{\smallskip}
Name & \object{CD\,Cet} & \\
GJ & \object{GJ 1057} & GJ91\\
Karmn & J03133+047 & Cab16 \\
$\alpha$ (J2000) & 03:13:22.92 & \textit{Gaia} DR2\\
$\delta$ (J2000) & +04:46:29.3 & \textit{Gaia} DR2\\
$G$ (mag) & $12.1232\pm0.0011$ & \textit{Gaia} DR2\\
$J$ (mag) & $8.775\pm0.020$ & 2MASS\\
Spectral type & M5.0\,V& PMSU\\
$d$ (pc) & $8.609\pm0.007$ & \textit{Gaia} DR2\\
$\mu_{\alpha}\cos{\delta}$ (mas\,a$^{-1}$) & $+1741.86 \pm0.17$ & \textit{Gaia} DR2\\
$\mu_{\alpha}$ (mas a$^{-1}$) & $+86.02\pm0.15$ & \textit{Gaia} DR2\\
$\gamma$ (km\,s$^{-1}$) & $+28.16\pm0.02$ & Laf20\\
$U$ (km\,s$^{-1}$) & $-60.32\pm0.05$ & This work \\
$V$ (km\,s$^{-1}$) & $-43.01\pm0.06$ & This work \\
$W$ (km\,s$^{-1}$) & $+19.27\pm0.07$ & This work \\
$T_{\rm eff}$ (K) & $3130\pm51$ & Pass19\\
$\log{g}$ (cgs) & 4.93$\pm0.04$ & Pass19\\
$\rm{[Fe/H]}$ (dex) & $0.13\pm0.16$ & Pass19\\
$L$ ($10^{-5}L_\odot$) & $293.4 \pm 5.3$ & Schw19\\
$M$ ($M_{\odot}$) & $0.161\pm0.010$ & Schw19\\
$R$ ($R_{\odot}$) & $0.175\pm0.006$ & Schw19\\
$v\sin{i}$ (km\,s$^{-1}$) & $\le 2.0$ & Rein18\\
$P_{\rm{rot}}$ (d) & 126.2 & New16\\
$\log{R'_{\rm{HK}}}$ & $-5.16$ & Boro18\\
Age (Ga) & 1--5 & This work \\ 
\noalign{\smallskip}
\hline
\end{tabular}
\tablebib{
   GJ91: \cite{GlJa91};
   2MASS: \cite{2MASS06};
   AF15: \cite{Alon15};
   Boro18: \cite{Activity_catalogue};
   Cab16: \cite{Carmencita};
   Laf20: \cite{Laf20};
   Gaia DR2: \cite{Gaia18b};
   New16: \cite{Newton_rotation};
   Pass19: \cite{Pass19});
   PMSU: \cite{1996AJ....112.2799H};
   Rein18: \cite{Reiners2018_324_stars};
   Schw19: \cite{Schw19}.
}
\end{table}

\object{CD\,Cet} has been observed by several spectroscopic and photometric instruments. Below, we shortly summarize the data available to us and analyzed in this work.

\subsection{Data}
\subsubsection{Precise radial velocities}

\paragraph{CARMENES.} 
Since the start of CARMENES operations in January 2016 we have collected $112$ VIS and $111$ NIR channel spectra within the CARMENES guaranteed time observations. 
The spectra typically reach S/N of $40$ in the VIS channel (at 0.66\,$\mu$m) and $150$ in the NIR channel (at 1.3\,$\mu$m) using exposure times of $30$\,min.
In the VIS channel, we discarded two spectra with low ${\rm S/N}<10$ and four without simultaneous FP drift measurement.
In the NIR channel, we discarded two spectra with ${\rm S/N}<10$, one without FP reference, and four spectra taken prior to the NIR channel stabilization (see Sect.~\ref{sec:NIR_performance}). 
This left us with 106 useful VIS and 105 NIR spectra.
We extracted, calibrated, and drift-corrected all of them with the CARMENES pipeline {\tt caracal} \citep{FOX, FP_Bauer, Caballero_data_flow}. 
RVs were derived from the useful orders selected in Sect.~\ref{sec:RV_computation} using {\serval} \citep{Zechmeister2018A&A...609A..12Z}. Subsequently, we corrected for systematic night-to-night offsets \citep{CARMENES_one_Trifon,HARPS_NZP, Lev_NZPs}. 
Our final RV time series exhibit an rms scatter of $5.2$\,m\,s$^{-1}$ with median errors of $1.7$\,m\,s$^{-1}$ in the VIS channel, and an rms scatter of $6.5$\,m\,s$^{-1}$ and a median error of $5.4$\,m\,s$^{-1}$ in the NIR channel. 
Therefore, both data sets indicate the presence of excess RV variability, either activity or planetary (or both).

\paragraph{ESPRESSO.} 
We acquired 17 additional spectra with the Echelle SPectrograph for Rocky Exoplanets and Stable Spectroscopic Observations  \citep[ESPRESSO;][]{Pepe2010SPIE.7735E..0FP}, the new ultra-stable spectrograph at the Very Large Telescope, between 20 and 26 September 2019. 
Typical S/N values are $21$ (at 0.66\,$\mu$m) for $10$\,min exposure times.
We used the ESO ESPRESSO pipeline data products and, for consistency, again derived RVs using {\serval}. The final {\serval} RV time series has an rms scatter of $5.6$\,m\,s$^{-1}$ with a median error of $0.8$\,m\,s$^{-1}$. RVs from the ESO pipeline exhibit a larger median uncertainty of $1.4$\,m\,s$^{-1}$, but both methods deliver mostly consistent RVs. We attribute the differences in RVs and precision to the fact that mid to late type M dwarfs have less isolated spectral features to perform cross-correlation, while the least-squares matching approach used by {\serval} is optimized for such conditions \citep{Guillem_TERRA}. 

\paragraph{HARPS.} There are nine publicly available spectra of \object{CD\,Cet} in the ESO archive\footnote{\url{http://archive.eso.org/eso/eso_archive_main.html}}, taken with the High Accuracy Radial velocity Planet Searcher \citep[HARPS,][]{Mayor2003Msngr.114...20M} between December 2003 and October 2010. 
Exposure times were $15$\,min in all cases. 
Although typical S/N values are $16$ (at 0.66\,$\mu$m), as for CARMENES, we discarded two spectra with ${\rm S/N} < 10$.
We computed the RVs of the remaining seven spectra using {\serval}, but the resulting time series yields an rms scatter of $29$\,m\,s$^{-1}$, which is five times larger than those in the CARMENES and ESPRESSO data sets.
Therefore, we did not use the HARPS data for the following analysis.

\subsubsection{Photometry}

\paragraph{MEarth.} 
The 8th data release\footnote{\url{https://www.cfa.harvard.edu/MEarth/DataDR8.html}} of the robotically-controlled telescope array MEarth tabulates photometric data of CD\,Cet collected between October 2008 and March 2018, spanning a total of 3456\,d (approximately 9.5 years). 
Over 114\,d in the 2010/2011 season, MEarth used an interference filter with which 93 measurements of CD\,Cet were collected. 
Because this data set covers only about one rotational period (following \citealt{Newton_rotation}) and used a different filter from the one used in the rest of the observations, we disregarded it for this work. 
The available photometric time series taken with the broad RG715 filter includes 11\,614 individual measurements, with an rms of $17$\,mmag and a median photometric precision of $4.3$\,mmag. 


\paragraph{TESS.} 
\object{CD\,Cet} was observed from 19 October 2018 to 14 November 2018 by the {\em Transiting Exoplanet Survey Satellite} \citep[{\em TESS},][]{Ricker2015} in Sector S04. The photometric time series consists of 18\,684 measurements with 2\,min cadence and a total time span of 25.95\,d. This is only a fraction of the rotational period derived from MEarth data, which made the {\em TESS} light curve of limited use for deriving (or confirming) the rotation period. 
However, both the photometric rms and photometric precision of a single measurement are typically $0.16$\,\%, that is, good enough to detect any transiting super-Earth companion to {CD\,Cet}.

\subsection{Analysis}\label{sec:analysis}

\subsubsection{Radial velocities}\label{sec:RV_ana}

\begin{figure}
    \includegraphics[width=\linewidth]{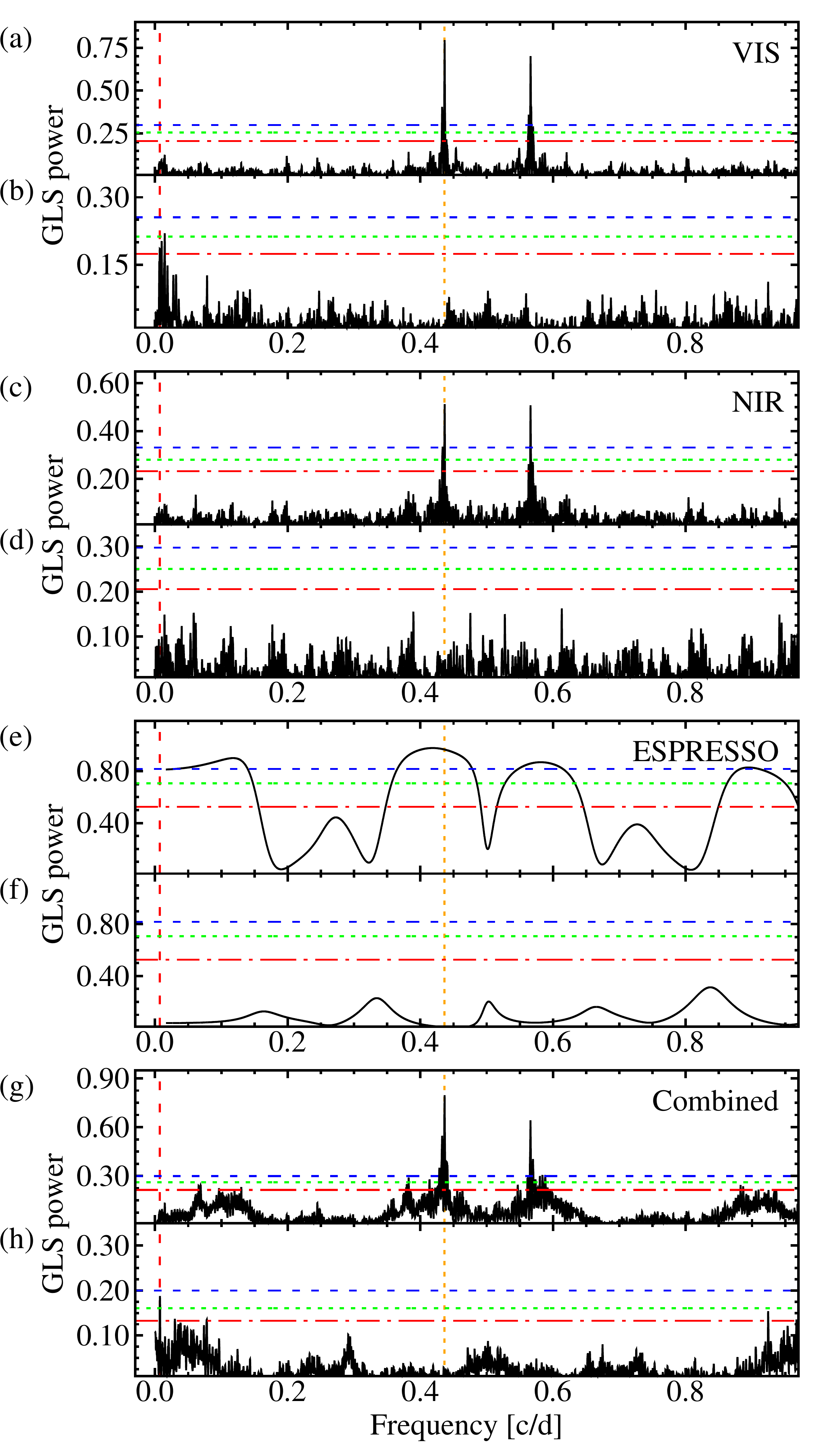} 
    \caption{GLS periodograms of the data from the VIS channel ($a$) and ($b$), the NIR channel ($c$) and ($d$), and ESPRESSO ($e$) and ($f$).
    Bottom panels always show the residual periodograms after subtracting the 2.3\,d signal. The 2.3\,d and the 134\,d signal are indicated by the vertical orange dotted and vertical red dashed line, respectively. 
    FAPs are indicated by horizontal lines: 10\,\% (dash-dotted red), 1\,\% (dotted green), 0.1\,\% (dashed blue). 
    }
    \label{fig:gls_rv} 
\end{figure}

We searched for periodic signals in the RV data by computing the generalized Lomb-Scargle (GLS) periodogram \citep{GLS}, as illustrated by Fig.~\ref{fig:gls_rv}. 
When we found a significant signal (i.e., false-alarm probability $\mathrm{FAP} < 0.1\,\%$), we made a Keplerian fit, removed it, and searched the residuals for more signals. 
The VIS (panel $a$) and NIR (panel $c$) channels both support the presence of a signal at $f_{\rm b}=0.435\,\mathrm{d}^{-1}$ ($\sim$2.3\,d) or $f'_{\rm b}=0.565\,\mathrm{d}^{-1}$ ($\sim$1.8\,d), which are related to each other via the one-day alias. 
Even though there are very few ESPRESSO RVs, we observed excess power at the same frequencies (panel $e$). 
Combining the data of CARMENES and ESPRESSO helped to resolve the daily alias issue (panel $g$), as both instruments are located at different geographical longitudes and the nightly duty cycle was thus improved. 
After the combination, the 2.3\,d signal is significantly preferred ($\mathrm{FAP_b} = 1.1 \times 10^{-15}$) over the 1.8\,d signal ($\mathrm{FAP'_b} = 5.6 \times 10^{-12}$).

After subtraction of the 2.3\,d signal, we find a tentative signal at $f=0.0147$\,d$^{-1}$ or 68\,d ($\mathrm{FAP_{VIS}} = 5\,\%$) in the VIS channel (panel $b$). 
This signal is at approximately half of the photometric rotation period derived by \cite{Newton_rotation}. 
There is no indication of the 68\,d period in the NIR channel data (panel $d$), which may be explained by a lower activity amplitude at redder wavelengths and by the higher instrumental jitter of the NIR channel data. 
The baseline of ESPRESSO (panel $f$) is too short to detect such long-term modulations. 
The combination of the three data sets (panel $h$) yielded a peak close to significance (FAP = 0.11\,\%) at $f=0.0074$\,d$^{-1}$ or 134\,d (about twice the period found with the VIS channel and very close to the photometric period).  
After removal of the 68--134\,d period, we did not find additional significant signals in the data.

\begin{table}
    \centering 
    \caption{Model evidence for different RV models.} 
    \label{tab:RV_signals} 
    \begin{tabular}{@{}lccc@{}} 
        \hline 
        \hline
        \noalign{\smallskip}
        Model      & $P$ (d)     & $\ln Z$       & $\Delta \ln Z$ \\
        \noalign{\smallskip}
        \hline 
        \noalign{\smallskip}
        Null & ... & $-743.80 \pm 0.12$ & 133.80 \\
        1 circ.\,orbit& 2.29 & $-630.33 \pm 0.16$ & 20.30  \\
        1 ecc.\,orbit & 2.29 & $-629.97 \pm 0.17$ & 19.97  \\
        2 ecc.\,orbits & 2.29; 139.2 & $-617.42 \pm 0.19$ & 7.42  \\
        1 circ.\,orbit\,+\,GP & 2.29; 152.5 & $-610.00 \pm 0.16$ & 0 \\
        1 ecc.\,orbit\,+\,GP & 2.29; 147.9 & $-608.73 \pm 0.17$ & --1.27 \\
        \noalign{\smallskip}
        \hline 
    \end{tabular} 
\end{table}

We modeled the combined RV data with {\tt juliet} \citep{juliet} to carry out a comparison between models containing one planet, two planets and one planet plus activity. 
Activity was modeled by a Gaussian process (GP) using a quasi-periodic kernel of the form:
\begin{equation}
    k(\tau) = A^{2} \exp\left(-\alpha \tau^2 - \Gamma \sin^2 \frac{\pi \tau}{P_{{\rm rot}}}\right),
    \label{eq:Qper}
\end{equation}
as implemented in {\tt george} \citep{Ambikasaran2015ITPAM..38..252A}. The parameter $A$ is the amplitude of the GP, $\alpha$ is the inverse of the exponential decay timescale, $\Gamma$ is the amplitude of the periodic (sine) part of the kernel, $P_{\rm rot}$ is its period, and $\tau=|t_i-t_j|$ is the time lag between any pair of RV measurements.

All modeling results are summarized in Table~\ref{tab:RV_signals}.  
As suggested by \cite{Bayesian_statistics_in_astrophysics}, we considered a model significantly better than another if the difference in Bayesian log-evidence exceeded $\Delta\ln Z$ = 5.  
The models using GPs to fit the second signal are, thus, preferred over the two-planet solution, which is a first indication that the 134\,d period is caused by activity.  
More evidence for this conclusion is presented in Sects.~\ref{sec:Act_ana} and~\ref{sec:Phot_ana}.  Allowing eccentric orbits in the fit did not significantly improve the Bayesian log-evidence ($\Delta\ln Z = -1.27$). 
Therefore, we adopted the circular orbit plus activity solution (``1 circ. orbit + GP'' in Table~\ref{tab:RV_signals}) for further analysis of the system.

\subsubsection{Chromatic and temporal coherence of the 2.3\texorpdfstring{\,}{~}d signal}\label{sec:coherence}

The RVs used in this paper were derived from observations over a wide range of different wavelengths, from the blue optical (ESPRESSO; 0.38\,$\mu$m) to the near infrared (CARMENES NIR; 1.7\,$\mu$m). 
We used the data of each instrument independently to test the chromaticity of the 2.3\,d signal. 
We removed the 134\,d signal for this analysis using the GP model of Sect.~\ref{sec:RV_ana} and fit a circular orbit for the 2.3\,d signal to the data. The Keplerian amplitudes found with each instrument are $K_{\rm ESPRESSO} = 6.93\pm0.41$\,m\,s$^{-1}$, $K_{\rm VIS} = 6.33\pm0.23$\,m\,s$^{-1}$, and $K_{\rm NIR}=6.59\pm0.84$\,m\,s$^{-1}$, which are all consistent within their error bars.

Most RV data for CD Cet were collected during the observing seasons 2018/19 and 2019/20. 
We also analyzed these seasons separately to probe period, phase, and amplitude changes of the 2.3\,d signal. 
Both data sets yield, however, consistent results, as shown in Table~\ref{tab:RV_seasons}. 
Chromatic and temporal coherence tests provide strong evidence for the origin of RV signals, while activity-induced signals are expected to show variations over wavelength and time. 
As this is not the case, the 2.3\,d signal is consistent with a planetary companion orbiting CD~Cet. 
However, the RV data gathered for this work were insufficient to test the 68--134\,d signal in the same way. 
Therefore, we resorted to spectral activity indicators (Sect.~\ref{sec:Act_ana}) and photometry (Sect.~\ref{sec:Phot_ana}) to resolve its origin.

\begin{table}
    \centering 
    \caption{Comparison of Keplerian parameters between the two main RV observing seasons.} 
    \label{tab:RV_seasons} 
    \begin{tabular}{@{}lccc@{}} 
        \hline 
        \hline 
        \noalign{\smallskip}
        Season & $P$ (d)     & $K$ (m\,s$^{-1}$) & $t_c-2458804$ (d) \\ 
        \noalign{\smallskip}
        \hline 
        \noalign{\smallskip}
        2018/19 & $2.2920 \pm 0.0022$ & $6.57\pm0.29$ & $0.530\pm0.052$\\
        2019/20 & $2.2917 \pm 0.0010$ & $6.40\pm0.30$ & $0.619\pm0.140$  \\
        \noalign{\smallskip}
        \hline 
    \end{tabular} 
\end{table}

\subsubsection{Activity indicators}\label{sec:Act_ana}

\begin{figure} 
     \includegraphics[width=\linewidth]{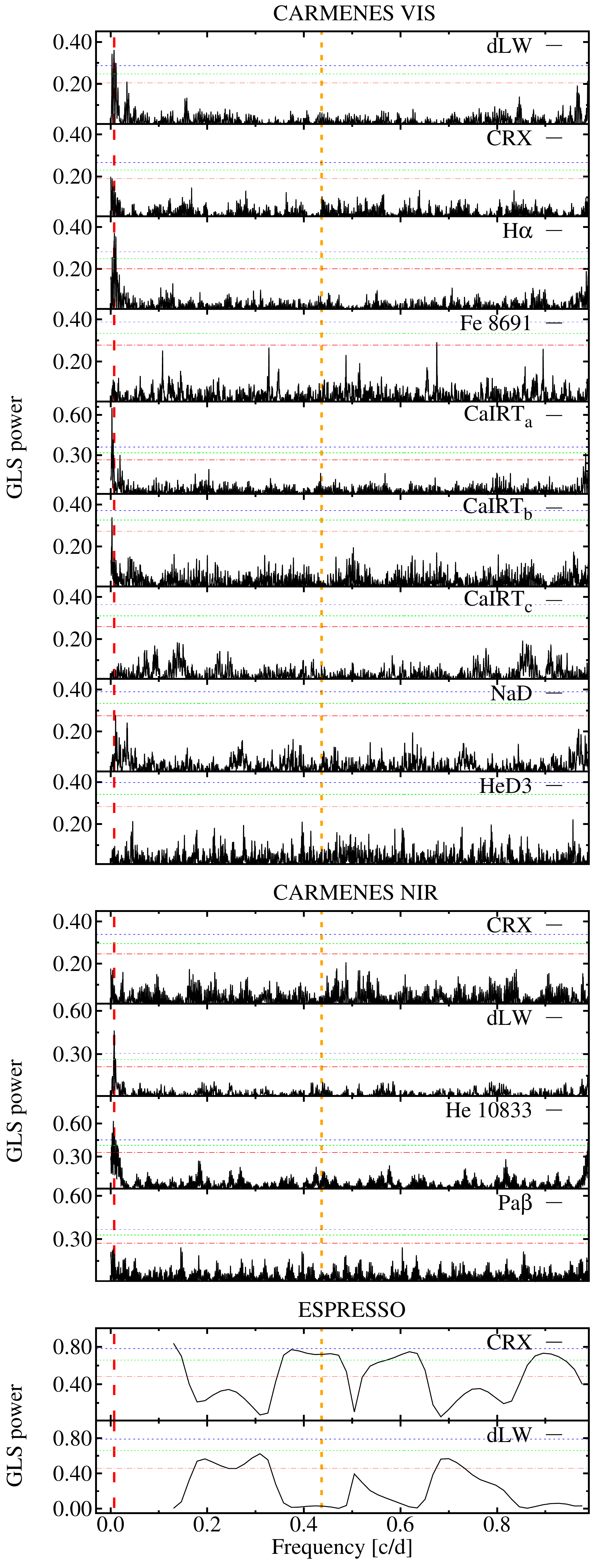} 
     \caption{GLS periodograms for various activity indicators. Horizontal FAP lines and vertical lines as in Fig.~\ref{fig:gls_rv}.
     }
     \label{fig:gls_activity} 
\end{figure}

We evaluated a number of activity indicators already used by us in a number of publications (e.g., \citealp{2018A&A...618A.115K, 2019A&A...622A.153N, 2019A&A...627A.116L, GJ3512, Teegarden}). 
They included the differential line width (dLW; similar to the FWHM of the cross-correlation function) and the chromatic index (CRX; wavelength dependence of an RV measurement) from {\serval}, as well as a number of line and band indices that track chromospheric activity from the VIS and NIR channel spectra. 
For a detailed explanation of all the activity indicators see \cite{Zechmeister2018A&A...609A..12Z} and \cite{Schoefer2019A&A...623A..44S}. 
For signal searches we computed the GLS periodograms displayed in Fig.~\ref{fig:gls_activity}. 

In two cases, namely the "a" component of the Ca~{\sc ii} infrared triplet (Ca~IRT$_{\rm a}$) covered by the VIS channel and the He~{\sc i} triplet at 1.083\,$\mu$m in the NIR channel, the strongest signals are related to a year or half-year period and were, therefore, disregarded. 
The Ca~IRT$_{\rm a}$ line in question is likely contaminated by a telluric line, while the He~{\sc i} triplet is located close to an OH sky emission line.  
Nonetheless, the dLW in the  VIS ($141$\,d) and the NIR channel ($134$\,d), as well as H$\alpha$ ($129$\,d), exhibit signals close to $134$\,d, where we found the second signal in our RV data.
This is yet another indication for stellar activity being the origin of this signal. 

In contrast, none of the spectral activity indicators shows any power at 2.3\,d. 
Combined with the information from chromatic and temporal coherence tests, we favor a planetary companion as explanation for the discovered RV signal.


\subsubsection{Photometry}\label{sec:Phot_ana}

\begin{figure}
    \includegraphics[width=\linewidth]{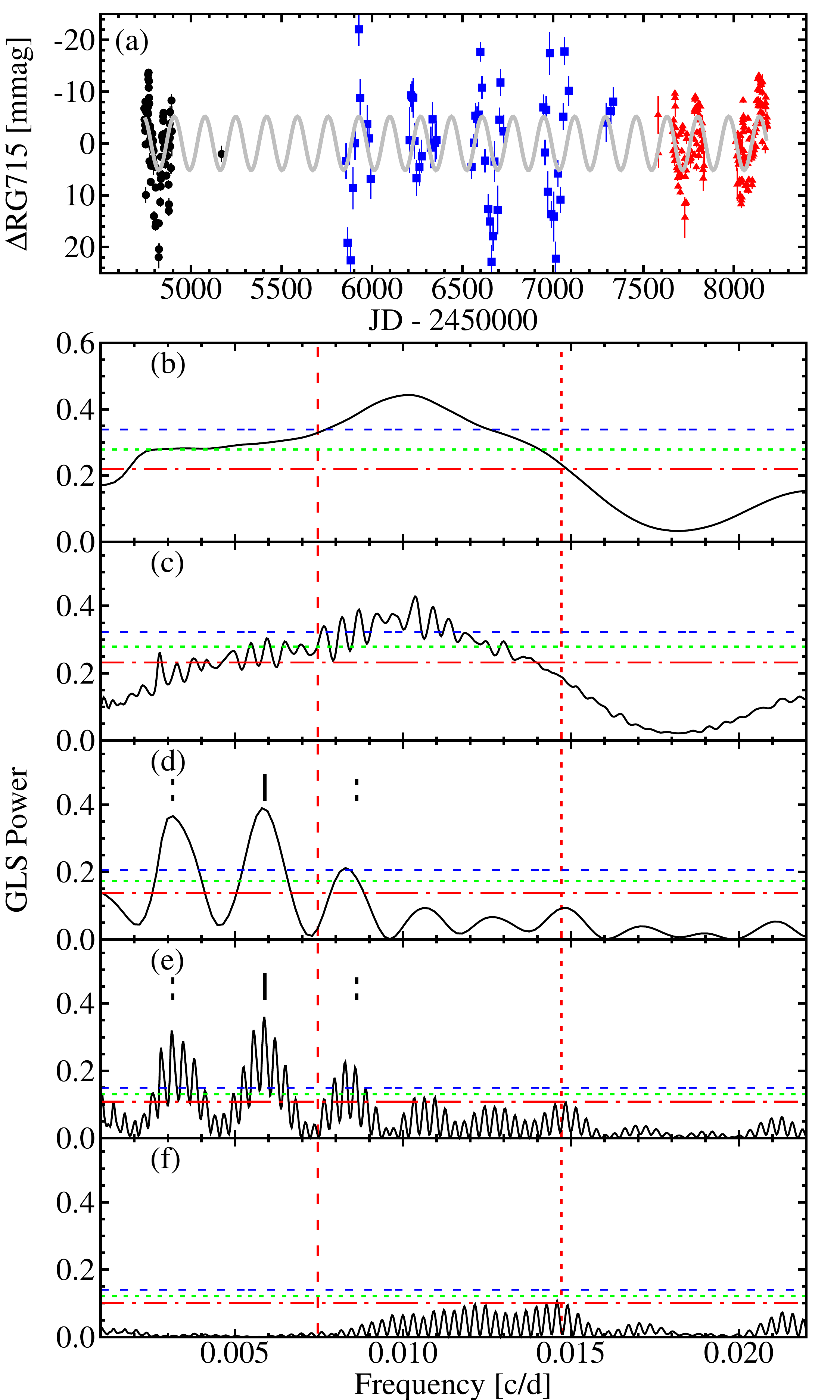} 
    \caption{MEarth photometry of CD\,Cet.  
    {\it Top panels}: ($a$) light curve used by \citet[black circles]{DA_rotation_periods}, \citet[adding blue squares]{Newton_rotation}, and this work (adding red triangles).  
    {\it Bottom panels:} ($b$) GLS periodograms of 
    data used by \cite{DA_rotation_periods}, ($c$) 
    data used by \cite{Newton_rotation}, ($d$) 
    2017/18 season data, ($e$) all public RG715 MEarth data, and ($f$) of their residuals after subtracting the 170\,d period.  
    Ticks in panels $d$ and $e$ mark a 170\,d period (solid black) and its one-year aliases (dashed black).  
    The vertical red dashed line marks the 134\,d period found in RV data and some spectral activity indicators, and the vertical red dotted line marks the 68\,d period found in VIS channel RV data.}
    \label{fig:gls_MEarth}
\end{figure}

Guided by the RV analysis results, we revisited the MEarth and {\em TESS} photometric data to confirm the rotation period of CD\,Cet and search for transits.
\cite{Newton_rotation} used the MEarth observations from 2008 to 2015 to derive a rotation period of about $126$\,d with a small amplitude of 6.2$\pm$2.2\,mmag. 
\cite{DA_rotation_periods}, using MEarth data spanning only the initial 1.2\,years of the observations (i.e., seasons 2008/09 and 2009/10), could not confirm it. 

In our analysis, we used all the RG715-filter CD~Cet data available in the 8th MEarth data release, which covers the years from 2008 to 2018 except for the interference-filter 2010/11 season gap.
We averaged multiple observations conducted within one night and derived a lightcurve consisting of 310 data points. In Fig.~\ref{fig:gls_MEarth} we present the whole light curve together with the GLS periodograms of the data used by \cite{DA_rotation_periods}, the data used by \cite{Newton_rotation}, the 2017/18 season data, which has the highest cadence and lowest rms, and the full RG715-filter MEarth dataset. 

As \cite{DA_rotation_periods}, we could not recover the 126.2\,d period found by \cite{Newton_rotation}. 
Furthermore, there is no indication of the tentative 68\,d and 134\,d periods found in the RVs and the spectral activity indicators H$\alpha$ and dLW of both channels. 
Although the MEarth photometry shows the presence of a long-period signal, its exact value can not be pinned down. 
When using the entire public MEarth dataset, we find the most significant (FAP$=2.2 \times 10^{-12}$\,\%) signal at about 170\,d. 
This period is dominated by the high cadence data of the 2017/18 season (panel $d$), while earlier sections of the MEarth data show very wide peaks at low frequencies (panels $b$ and $c$). 

\begin{table}
    \caption{\label{tab:photometry_GP} Priors and posteriors of our GP fit to MEarth photometry.}
    \centering
    \renewcommand{\arraystretch}{1.3}
    \begin{tabular}{@{}llll@{}}
        \hline
        \hline
        \noalign{\smallskip}
        Parameter & Prior value & Posterior & Unit \\
        \noalign{\smallskip}
        \hline
        \noalign{\smallskip}
        \multicolumn{4}{c}{\em Model 1: Quasi-periodic kernel} \\
        $A_{\Delta\,\mathrm{RG715}}$ & $\PriU(0,300)$ & $7.2_{-0.7}^{+0.8}$ & mmag \\
        $\alpha$ & $\PriJ(10^{-5},1)$ & $5.9_{-3.8}^{+77.4}$ & $10^{-5}$\,d$^{-2}$\\
        $\Gamma_{\Delta\,\mathrm{RG715}}$ & $\PriU(0,300)$ & $0.012_{-0.005}^{+0.010}$ & mmag \\
        $P_{\rm rot}$ & $\PriU(100,200)$ & $170_{-38}^{+19}$ & d \\
        $\sigma_{\mathrm{MEarth}}$ & $\PriU(0.1,300)$ & $3.7_{-0.3}^{+0.3}$ & mmag \\
        $\mu_{\mathrm{MEarth}}$ & $\PriU(-200,200)$ & $-0.8_{-0.9}^{+1.0}$ & $10^{-6}$\,mmag \\
        \noalign{\smallskip}
        \multicolumn{4}{c}{\em Model 2: Matern kernel} \\
        $A_{\Delta\,\mathrm{RG715}}$ & $\PriU(0,300)$ & $7.4_{-0.7}^{+0.9}$ & mmag \\
        $\rho$ & $\PriJ(10^{-5},1000)$ & $13.8_{-2.7}^{+3.1}$ & $\mathrm{d}$\\
        $\sigma_{\mathrm{MEarth}}$ & $\PriU(0.1,300)$ & $3.4_{-0.2}^{+0.2}$ & mmag \\
        $\mu_{\mathrm{MEarth}}$ & $\PriU(-200,200)$ & $-0.4_{+1.2}^{-1.2}$ & $10^{-6}$\,mmag \\
        \noalign{\smallskip}
        \hline
    \end{tabular}
    \tablefoot{
    $\sigma$ is the instrumental jitter, $\mu$ is the fit offset, and $\PriU$ and $\PriJ$ stand for uniform and log-uniform distributions, respectively \citep{Je46}.
    }
\end{table}

Because we observed phase and amplitude changes in the photometry (see panel $a$ of Fig.~\ref{fig:gls_MEarth}), we fit the MEarth data with a GP.  Table~\ref{tab:photometry_GP} summarizes the two GP models.
When fitting a GP with the quasi-periodic kernel of Eq.~(\ref{eq:Qper}), we found a very low amplitude for the periodic term, $\Gamma = 0.012$\,mmag. 
This is a hint that the photometric variability over a decade is not dominated by rotation, but by spot evolution (represented by the exponential decay term $\alpha$ in Eq.~(\ref{eq:Qper})). 
Hence, we reduced the number of parameters and used a stationary Matern kernel (with amplitude $A$ and a correlation length $\rho$ as parameters) to just fit for spot evolution with a GP.
The Matern kernel gave a significantly higher Bayesian log-evidence than the quasi-periodic kernel ($\Delta\ln Z=13.27$).
As a result, spot evolution explains the frequency changes observed between different sections of MEarth photometry. 
We could also expect to find period differences between photometry on the one hand and RVs and the spectral activity indicators on the other hand, as they span different epochs.   

\begin{figure} 
    \includegraphics[width=\linewidth]{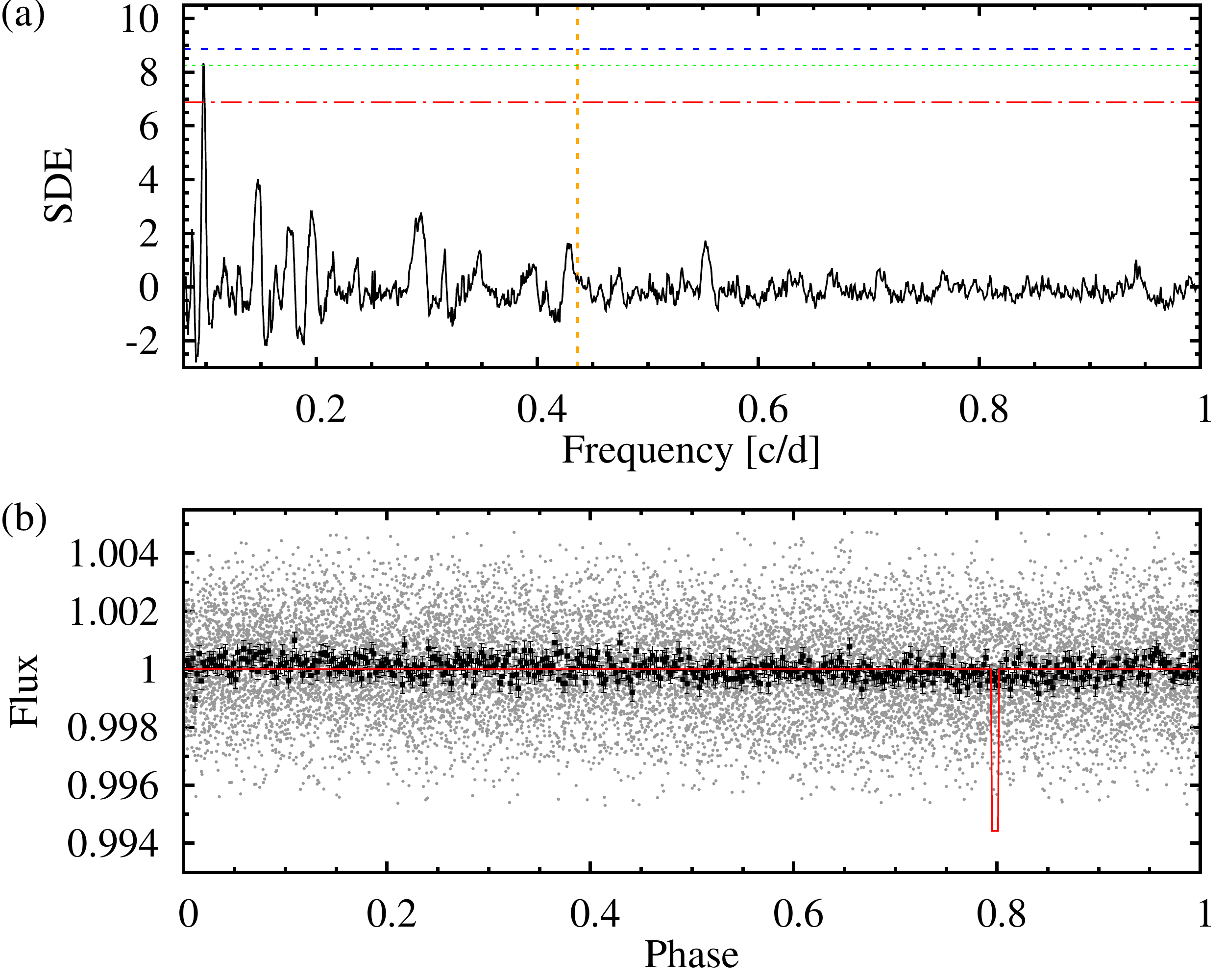}
    \caption{Transit search in the {\tess} lightcurve. 
    {\it Top:} signal detection efficiency (SDE) as a function of frequency using the Transit Least Squares algorithm. Horizontal lines indicates false alarm probabilities of 10\,\% (red), 1\,\% (green), and 0.1\,\% (blue). 
    Vertical orange dotted line marks the planetary period $P_{\rm b}$ = 2.2907\,d present in the RVs.
    {\it Bottom:} {\tess} data (gray: two minute cadence, black: binned to 500 phase points) folded to $P_{\rm b}=2.2907$\,d guided by the RV solution along with the expected transit signal (red solid line).
    }
    \label{fig:transit_search} 
\end{figure}

With an orbital period of 2.3\,d, the transit probability of CD\,Cet\,b is 4.4\,\%, and we would expect ten transits during the 25\,days of observation. 
No alert was issued for this target after the end of observations of {\em TESS} Sector 04 in November 2018.
We re-investigated the light curve with the Transit Least Squares algorithm \citep{TLS}. 
As expected, we found no indication of a transit, as shown in Fig.~\ref{fig:transit_search}. 
We used the winning RV model of Sect.~\ref{sec:RV_ana} to estimate the transit center in Fig.~\ref{fig:transit_search}. With an expected transit depth of $0.56\,\%$ (assuming a rocky and silicate composition; \citealp{Zeng_2019}), CD\,Cet\,b would have been easily detected by {\tess} (rms $=0.16$\,\%). Thus, we concluded that CD\,Cet\,b is not transiting.

\subsubsection{Combined}\label{sec:comb_ana}

\begin{figure} 
    \includegraphics[width=\linewidth]{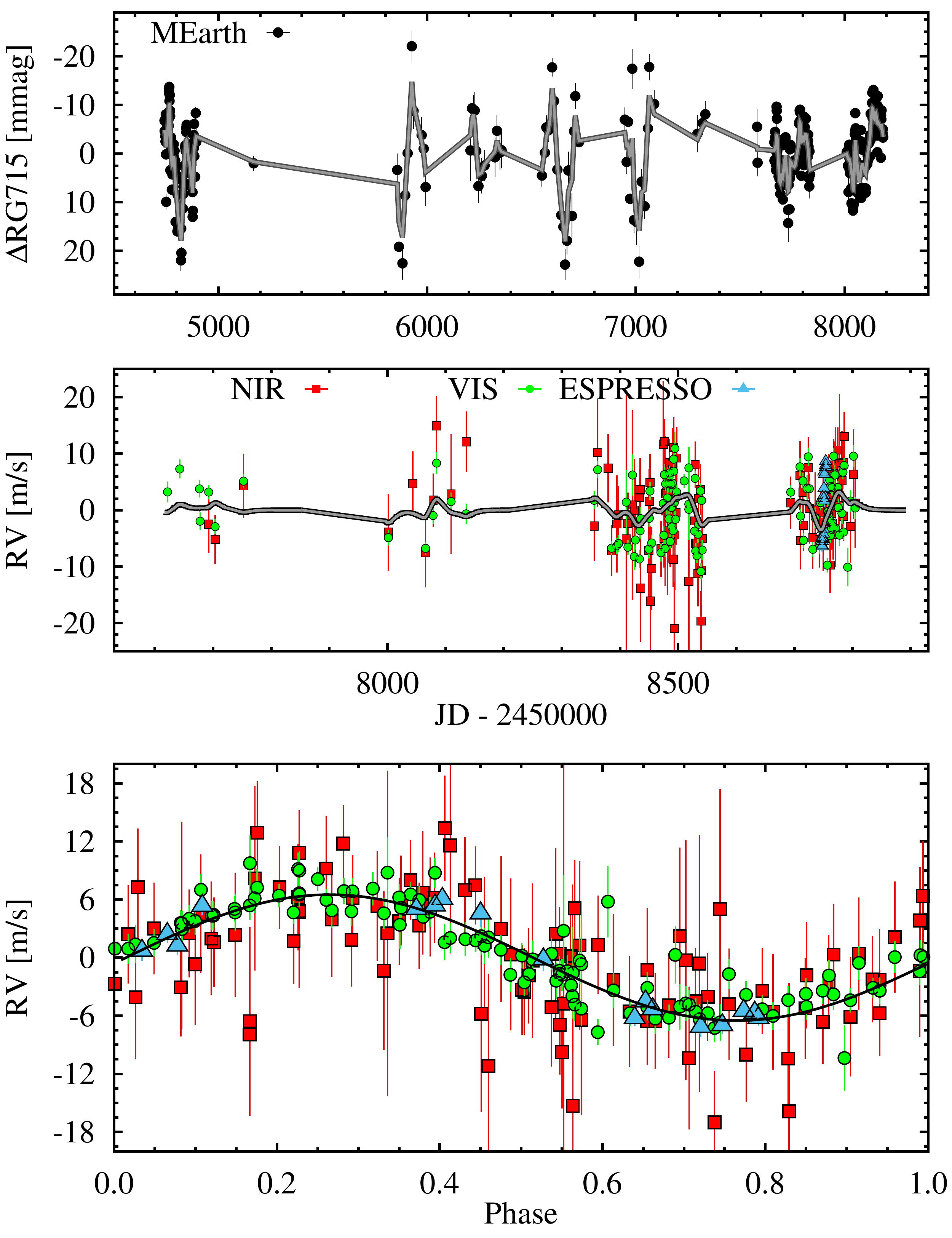} 
    \caption{Combined analysis of photometry and RV. 
    {\it Top panel:} MEarth photometry (black points) and GP fit (gray line).  
    {\it Middle panel:} RV time series of the VIS channel (green points), the NIR channel (red squares), ESPRESSO (blue triangles), and GP fit with Matern kernel (gray line).  {\it Bottom panel:} activity-subtracted RVs folded to the Keplerian period $P_{\rm b}=2.2907$\,d and best-fit circular orbit (black line).
    }
    \label{fig:GP_phot_rv} 
\end{figure}

In order to derive the planetary parameters of CD\,Cet\,b, we took into account the activity signal. 
We combined photometric and RV information in a joint fit of CARMENES, ESPRESSO, and MEarth data. 
Activity was fit by a GP, which was constrained by RVs and photometry simultaneously. 
For completeness, we tested again quasi-periodic and Matern kernels, and concluded that indeed the Matern kernel was the best choice for the combined data set as well ($\Delta\ln Z=11.74$).
Because the Bayesian model comparison in Sect.~\ref{sec:RV_ana} favored assuming zero eccentricity, we fit a circular orbit for CD\,Cet\,b. 
All model parameters are listed in Table~\ref{tab:RV+Phot_analysis}, while the resulting fits to the RVs and the photometric data can be seen in Fig.~\ref{fig:GP_phot_rv}.
When subtracting the model from the data, we obtained a weighted rms consistent with the respective error bars in each instrument: $1.55$\,m\,s$^{-1}$ in the VIS channel, $4.66$\,m\,s$^{-1}$ in the NIR channel, and $1.13$\,m\,s$^{-1}$ for ESPRESSO. Assuming a variety of Bond albedos found among the rocky Solar System bodies, $A_0 \approx 0$ (P-type asteroids), $A_\oplus \approx 0.30$, or $A_{\rm Venus} \approx 0.76$, results in a wide range of equilibrium temperatures from $T_{{\rm eq},0} = 464\pm16$\,K to $T_{{\rm eq},\oplus} = 424\pm16$\,K down to $T_{\rm eq,Venus} = 325\pm16$\,K.
We conclude that CD\,Cet is orbited by a temperate super-Earth ($3.95\,M_\oplus$) on a 2.29\,d orbit.

\begin{figure} 
    \includegraphics[width=\linewidth]{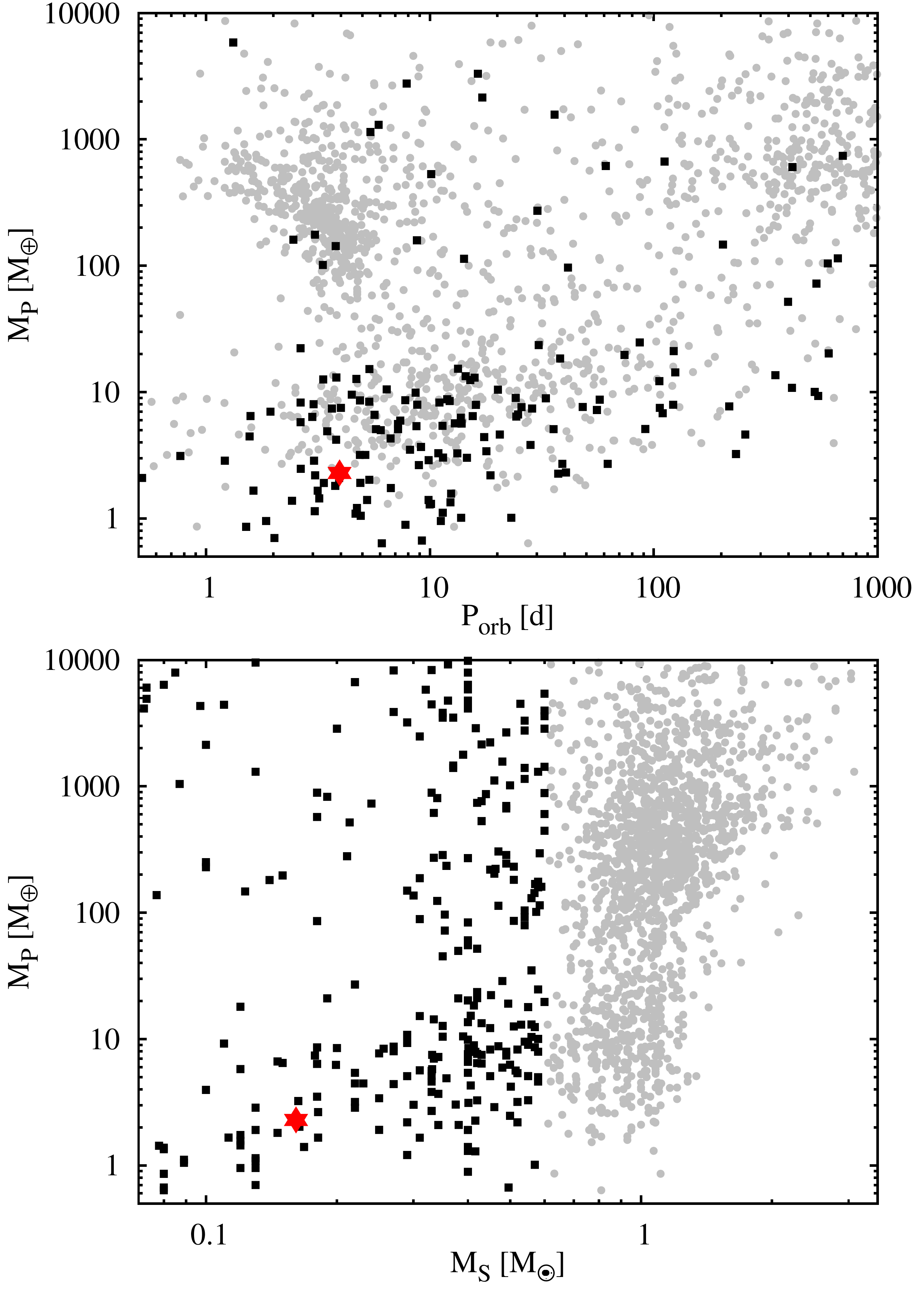} 
    \caption{{\it Top panel:} orbital period versus planetary mass for known exoplanets orbiting stars $>0.6\,M_{\odot}$ (gray points), $\le 0.6\,M_{\odot}$ (black squares), and CD Cet b (red star). {\it Bottom panel:} stellar mass versus planetary mass plot. Colors and symbols as above.}
    \label{fig:properties} 
\end{figure}

To compare CD\,Cet\,b to the bulk of known exoplanets\footnote{taken from \url{exoplanet.eu}}, we show the period versus planet mass and stellar mass versus planet mass plots in Fig.~\ref{fig:properties}. CD\,Cet\,b lines up well with other discoveries; it belongs to a population of super-Earths commonly found around M~dwarfs. The discovery adds to 42 other known exoplanets in the regime of low-mass exoplanets ($<5 M_{\oplus}$) orbiting mid- to late-type M~dwarfs ($<0.3 M_{\odot}$).

\begin{table}
    \caption{\label{tab:RV+Phot_analysis} Prior and posterior values of the combined planet plus activity fit to RVs and MEarth photometry and derived planetary parameters.}
    \centering
    \setlength{\tabcolsep}{5.0pt}
    \renewcommand{\arraystretch}{1.3}
    \begin{tabular}{@{}lccl@{}}
        \hline
        \hline
        \noalign{\smallskip}
        Parameter & Prior & Posterior & Unit \\
        \noalign{\smallskip}
        \hline
        \noalign{\smallskip}
        \multicolumn{4}{c}{\em Planet}\\
        $P_{\rm b}$ & $\PriU(2.0,2.5)$ & $2.29070_{-0.00012}^{+0.00012}$ & d \\
        $K_{\rm b}$ & $\PriU(0,10)$ & $6.51_{-0.23}^{+0.22}$ & m\,s$^{-1}$ \\
        $t_{c,{\rm b}}-2450000$ & $\PriU(8802.7,8805.0)$ & $8804.489_{-0.018}^{+0.017}$ & d \\
        \noalign{\smallskip}
        \multicolumn{4}{c}{\em Matern kernel}\\
        $A_{\Delta\,\mathrm{RG715}}$ & $\PriU(0,300)$ & $7.46_{-0.72}^{+0.84}$ & mmag \\
        $A_{\mathrm{RV}}$ & $\PriU(0,40)$ & $2.05_{-0.34}^{+0.40}$ & m\,s$^{-1}$ \\
        $\rho$ & $\PriJ(10^{-5},1000)$ & $14.6_{-2.4}^{+2.9}$ & $\mathrm{d}$\\
        \noalign{\smallskip}
        \multicolumn{4}{c}{\em Instruments} \\
        $\sigma_{\mathrm{MEarth}}$ & $\PriU(0.1,300)$ & $3.46_{-0.21}^{+0.22}$ & mmag \\
        $\mu_{\mathrm{MEarth}}$ & $\PriU(-200,200)$ & $-0.4_{-1.2}^{+1.2} \times 10^{-6}$ & mmag \\
        $\sigma_{\text{C.VIS}}$ & $\PriJ(0.01,15)$ & $0.84_{-0.21}^{+0.27}$ & m\,s$^{-1}$ \\
        $\mu_{\text{C.VIS}}$ & $\PriU(-10,10)$ & $-0.74_{-0.56}^{+0.54}$ & m\,s$^{-1}$ \\
        $\sigma_{\text{C.NIR}}$ & $\PriJ(0.01,20)$ & $0.97_{-0.34}^{+0.57}$ & m\,s$^{-1}$ \\
        $\mu_{\text{C.NIR}}$ & $\PriU(-20,20)$ & $0.08_{-0.79}^{-0.76}$ & m\,s$^{-1}$ \\
        $\sigma_{\mathrm{ESPRESSO}}$ & $\PriJ(0.01,15)$ & $0.91_{-0.34}^{+0.41}$ & m\,s$^{-1}$ \\
        $\mu_{\mathrm{ESPRESSO}}$ & $\PriU(-20,20)$ & $4.4_{+1.1}^{-1.1}$ & m\,s$^{-1}$ \\
        \noalign{\smallskip}
        \multicolumn{4}{c}{\em Derived}\\
        $a_{\rm b}$ & & $0.0185_{-0.0013}^{+0.0013}$ & au\\
        $m_{\rm b} \sin{i}$ & & $3.95_{-0.43}^{+0.42}$ & $M_\oplus$ \\
        $i_{\rm b}$ & &$<87.48_{-0.28}^{+0.25}$ & deg \\
        $S_{\rm b}$ & & $8.6_{-2.4}^{+2.4}$ & $S_{\oplus}$ \\
        $T_{\rm{eq,b}}$ & & $464_{-18}^{+18}$  & K \\
        \noalign{\smallskip}
        \hline 
\end{tabular}
\end{table}

\section{Summary and discussion}\label{sec:discussion}

In this work we presented an overview of how the optical and near-infrared dual-channel CARMENES spectrograph can be used to deliver high-precision RVs. Scientists who want to use CARMENES in the future should understand the instrument and its performance on the level that was discussed in this paper. 
This will be of importance for the community, as a considerable amount of observing time is available through open calls for proposals. Moreover, there are many other similar instruments that achieved first light after CARMENES and that may benefit from the concepts discussed in this work. 

Currently, CARMENES VIS is capable of routinely measuring RVs with a median precision of 1.2\,m\,s$^{-1}$. The main instrumental limitations are currently imperfections in the calibration, which may be related to the sensitivity of the instrument to temperature changes in the climatic rooms, to drifts of the FP etalons, and to inconsistencies in the wavelength solutions computed every night. Improvements through higher thermal stability, second-generation etalons, and improved procedures for the calibration and wavelength solution would further reduce the instrumental errors.

On the other hand, CARMENES NIR delivers a median RV precision of 3.7\,m\,s$^{-1}$ (around three times that of CARMENES VIS). The key to improving the performance of this channel lies in optimizing its active thermal stabilization mechanism. This has already been partly achieved during the survey, and further options are currently under investigation.

We showed that carefully avoiding spectral regions with low RV content and strong telluric contamination or regions with increased detector effects can significantly improve the RV performance. For CARMENES NIR and with the currently implemented strategy of masking telluric contamination, optimal performance was reached if only half of the covered spectral range was used to measure RVs. However, in the future, efforts for telluric correction may result in larger useful spectral regions. The rejection strategy led to an average improvement of 2\,m\,s$^{-1}$ in the rms of our NIR channel RV curves. 
On the whole, we showed that the RV rms scatter in our M dwarf sample is consistent with expectations from the combined effects of photon noise, instrumental performance, and stellar activity.

As an example, we presented the case of CD\,Cet. This M5.0\,V star is a typical target for which the instrument was designed. 
With about 100 spectra taken, we discovered a highly significant signal of amplitude $K=6.5$\,m\,s$^{-1}$ independently in both CARMENES channels. The facts that the amplitude was consistent in the CARMENES VIS and NIR channels, and that none of our activity indicators showed a $2.3$\,d period, are strong indicators for the planetary (super-Earth) origin of the signal. 
The signal was confirmed by 17 ESPRESSO measurements complementing CARMENES at bluer wavelengths.

The ability to investigate the consistency of an RV signal across large wavelength ranges and, thereby, differentiate between signals caused by stellar activity from those of Keplerian nature, is one of the main strengths of CARMENES. 
The use of chromaticity at wavelengths longer than 1\,$\mu$m is, however, hampered by the NIR channel photon limit, which is typically on the order of 5--7\,m\,s$^{-1}$ (assuming S/N\,=\,150). 
A second, low-amplitude ($\approx2$\,m\,s$^{-1}$) RV signal in CD\,Cet was, hence, only picked up in the VIS channel. However, this signal in the range of 130--140 days had counterparts in H$\alpha$ and the dLW, measured in both CARMENES VIS and NIR. CD\,Cet appears to be a relatively quiet star, which is also reflected by a low photometric variability of about 7\,mmag, again found over long time scales. 
The discovery of CD\,Cet\,b is, thus, an excellent example of the capabilities of CARMENES, which was designed for discovering and extending our statistical knowledge of planets orbiting mid-M dwarfs, and for discriminating their Keplerian signatures from activity-induced RV variations. Currently, planned upgrades have the potential of reducing the instrumental noise in both channels of CARMENES even further.




\begin{acknowledgements} 
We thank the referee for her/his suggestions that have certainly improved this manuscript.
We thank the whole team at CAHA, because their work during day and night time makes our science with CARMENES possible: 
Jes\'us Aceituno, 
Bel\'en Arroyo, 
Antonio Bar\'on Carre\~no, 
Daniel Ben\'itez, 
Gilles Bergond, 
Enrique de Guindos, 
Enrique de Juan, 
Alba Fern\'andez, 
Roberto Fern\'andez Gonz\'alez, 
David Galad\'i-Enriquez, 
Eulalia Gallego, 
Joaqu\'in García de la Fuente, 
Vicente G\'omez Galera, 
Jos\'e G\'ongora Rueda, 
Ana Guijarro, 
Jens Helmling, 
Israel Hermelo, 
Luis Hern\'andez Casta\~no, 
Francisco Hern\'andez Hernando, 
Juan F. L\'opez Salas, 
H\'ector Mag\'an Medinabeitia, 
Julio Mar\'n Molina, 
Pablo Mart\'in, 
David Maroto Fern\'andez, 
Francisco M\'arquez Sánchez, 
Santos Pedraz, 
Rub\'en P. Hedrosa, 
Miguel \'A. Pe\~nalver, 
Manuel Pineda Puente, 
Santiago Reinhart Franca,
and
Jos\'e I. Vico Linares.
CARMENES is an instrument for the Centro Astron\'omico Hispano-Alem\'an (CAHA) at Calar Alto (Almer\'{\i}a, Spain), operated jointly by the Junta de Andaluc\'ia and the Instituto de Astrof\'isica de Andaluc\'ia (CSIC).
CARMENES was funded by the German Max-Planck-Gesellschaft (MPG), 
the Spanish Consejo Superior de Investigaciones Cient\'{\i}ficas (CSIC),
the European Union through FEDER/ERF FICTS-2011-02 funds, 
and the members of the CARMENES Consortium 
  (Max-Planck-Institut f\"ur Astronomie,
  Instituto de Astrof\'{\i}sica de Andaluc\'{\i}a,
  Landessternwarte K\"onigstuhl,
  Institut de Ci\`encies de l'Espai,
  Institut f\"ur Astrophysik G\"ottingen,
  Universidad Complutense de Madrid,
  Th\"uringer Landessternwarte Tautenburg,
  Instituto de Astrof\'{\i}sica de Canarias,
  Hamburger Sternwarte,
  Centro de Astrobiolog\'{\i}a and
  Centro Astron\'omico Hispano-Alem\'an), 
with additional contributions by the Spanish Ministry of Economy, 
the German Science Foundation through the Major Research Instrumentation Program and DFG Research Unit FOR2544 ``Blue Planets around Red Stars'', 
the Klaus Tschira Stiftung, 
the states of Baden-W\"urttemberg and Niedersachsen, and by the Junta de Andaluc\'{\i}a.
Based on data from the CARMENES data archive at CAB (INTA-CSIC).
We acknowledge financial support from
the European Research Council under the Horizon 2020 Framework Program via the ERC Advanced Grant Origins 83~24~28,
the Deutsche Forschungsgemeinschaft through project RE~1664/14-1,
the Agencia Estatal de Investigaci\'on of the Ministerio de Ciencia, Innovaci\'on y Universidades and the European FEDER/ERF funds through projects    
  AYA2018-84089, 		    
  ESP2016-80435-C2-1-R,		
  AYA2016-79425-C3-1/2/3-P,	
  AYA2015-69350-C3-2-P,		
the Centre of Excellence ``Severo Ochoa'' and ``Mar\'ia de Maeztu'' awards to the Instituto de Astrof\'isica de Canarias (SEV-2015-0548), Instituto de Astrof\'isica de Andaluc\'ia (SEV-2017-0709), and Centro de Astrobiolog\'ia (MDM-2017-0737), 
  and the Generalitat de Catalunya/CERCA program.

\end{acknowledgements} 


\bibliographystyle{aa} 


\begin{appendix}

\section{Cornerplot and RV table}

\begin{figure*} 
    \includegraphics[width=\linewidth]{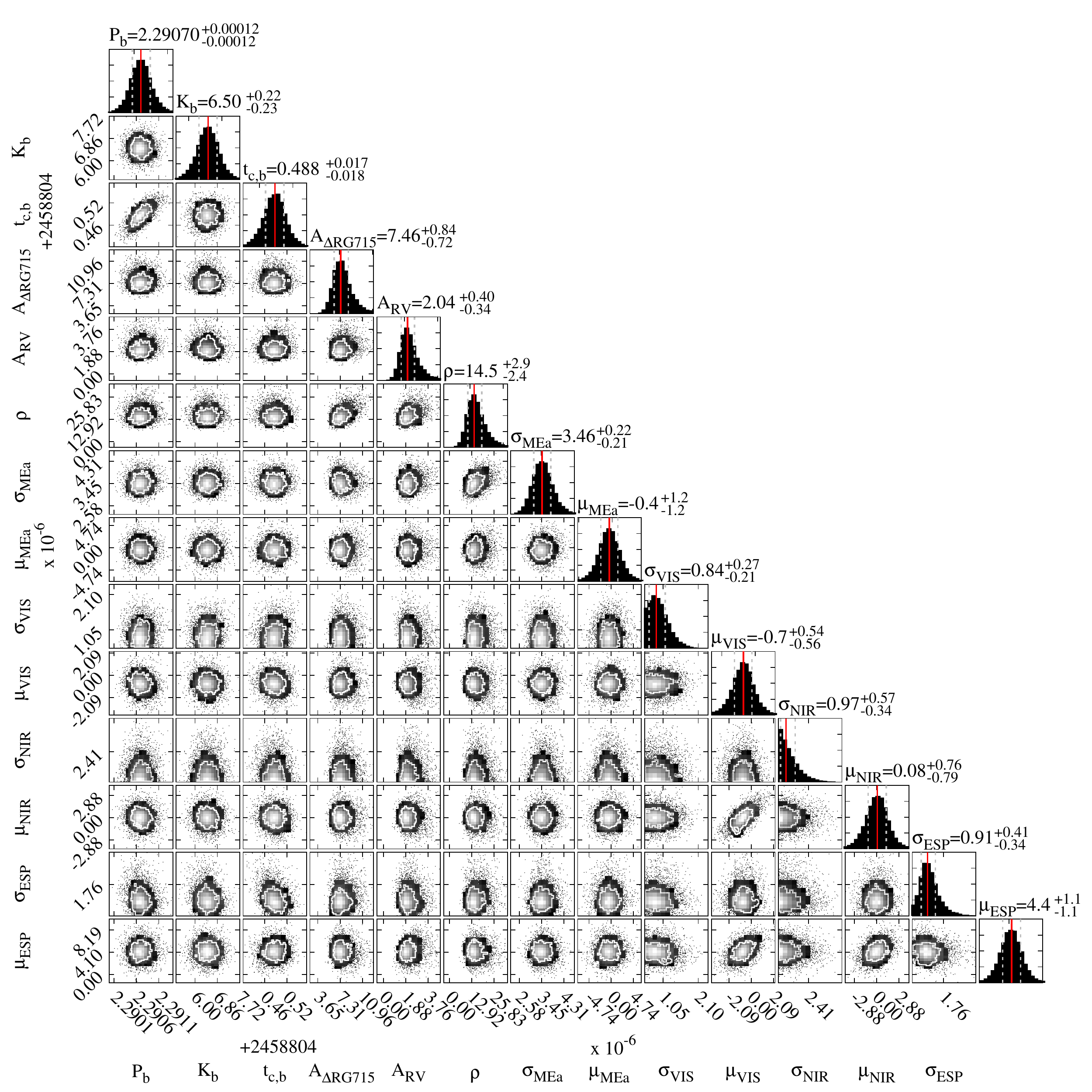} 
    \caption{Sample distribution for the combined fit of RVs and MEarth photometry (Table~\ref{tab:RV+Phot_analysis}). The vertical line in each histogram indicates the median (50\,\% quantile, solid red) and the 1\,$\sigma$ uncertainties (16\,\% and 84\,\% quantiles, dashed gray).
    }
    \label{fig:corner} 
\end{figure*}

\begin{center}
\onecolumn
\begin{longtable}{@{}cccl@{}}
    \caption{\label{tab:RVs} Radial velocities used in this paper.}\\
        \hline
        \hline
        \noalign{\smallskip}
        BJD & RV & $\delta$RV [m\,s$^{-1}$] & Instrument \\
            & [m\,s$^{-1}$] & [m\,s$^{-1}$] & \\
        \noalign{\smallskip}
        \hline
        \endfirsthead
                \caption{Radial velocities used in this paper (cont.)}\\
        \hline
        \hline
        BJD & RV & $\delta$RV [m\,s$^{-1}$] & Instrument \\
            & [m\,s$^{-1}$] & [m\,s$^{-1}$] & \\
        \hline
        \endhead
        \hline
        \endfoot
        \noalign{\smallskip}
2457621.67888	&	3.22	&	1.86	&	CARM-VIS	\\
2457642.68307	&	7.28	&	1.69	&	CARM-VIS	\\
2457676.59512	&	3.76	&	1.49	&	CARM-VIS	\\
2457677.59061	&	-1.97	&	1.59	&	CARM-VIS	\\
2457692.53769	&	3.19	&	1.28	&	CARM-VIS	\\
2457703.50752	&	-2.94	&	1.31	&	CARM-VIS	\\
2457752.42983	&	5.13	&	1.28	&	CARM-VIS	\\
2458001.57254	&	-4.88	&	1.36	&	CARM-VIS	\\
2458065.51706	&	-6.79	&	1.19	&	CARM-VIS	\\
2458078.44269	&	-0.98	&	1.48	&	CARM-VIS	\\
2458084.46363	&	8.32	&	1.91	&	CARM-VIS	\\
2458109.44990	&	1.50	&	1.61	&	CARM-VIS	\\
2458135.38699	&	-0.66	&	1.83	&	CARM-VIS	\\
2458355.59568	&	1.74	&	1.62	&	CARM-VIS	\\
2458361.63224	&	7.15	&	1.48	&	CARM-VIS	\\
2458385.64435	&	-6.75	&	1.29	&	CARM-VIS	\\
2458394.59634	&	-6.33	&	1.42	&	CARM-VIS	\\
2458397.62191	&	-5.93	&	1.57	&	CARM-VIS	\\
2458410.60644	&	1.40	&	5.71	&	CARM-VIS	\\
2458413.58058	&	-6.57	&	1.29	&	CARM-VIS	\\
2458421.56441	&	6.21	&	3.66	&	CARM-VIS	\\
2458424.67833	&	-8.26	&	2.33	&	CARM-VIS	\\
2458426.49175	&	-5.30	&	1.70	&	CARM-VIS	\\
2458433.54508	&	-8.66	&	1.62	&	CARM-VIS	\\
2458434.52675	&	-3.67	&	1.42	&	CARM-VIS	\\
2458435.59257	&	-1.52	&	2.06	&	CARM-VIS	\\
2458449.59827	&	-6.92	&	1.63	&	CARM-VIS	\\
2458450.48081	&	-1.45	&	1.61	&	CARM-VIS	\\
2458451.46339	&	3.37	&	1.54	&	CARM-VIS	\\
2458454.48322	&	-5.84	&	1.46	&	CARM-VIS	\\
2458470.42316	&	-7.54	&	1.57	&	CARM-VIS	\\
2458474.42725	&	1.16	&	1.51	&	CARM-VIS	\\
2458475.42822	&	-4.47	&	1.56	&	CARM-VIS	\\
2458476.41889	&	6.27	&	1.33	&	CARM-VIS	\\
2458477.41990	&	-6.84	&	1.49	&	CARM-VIS	\\
2458478.40155	&	4.64	&	1.24	&	CARM-VIS	\\
2458480.39332	&	0.75	&	1.52	&	CARM-VIS	\\
2458481.38642	&	2.13	&	1.88	&	CARM-VIS	\\
2458483.40355	&	4.68	&	1.54	&	CARM-VIS	\\
2458484.37463	&	-1.56	&	1.51	&	CARM-VIS	\\
2458485.39910	&	6.57	&	1.16	&	CARM-VIS	\\
2458486.38676	&	-5.57	&	1.26	&	CARM-VIS	\\
2458487.50047	&	4.54	&	1.45	&	CARM-VIS	\\
2458488.37346	&	0.48	&	1.23	&	CARM-VIS	\\
2458489.38051	&	-3.03	&	1.09	&	CARM-VIS	\\
2458490.38613	&	4.77	&	1.44	&	CARM-VIS	\\
2458491.41350	&	-3.64	&	1.45	&	CARM-VIS	\\
2458492.38027	&	9.02	&	1.17	&	CARM-VIS	\\
2458492.40442	&	6.87	&	1.11	&	CARM-VIS	\\
2458493.38641	&	1.34	&	2.92	&	CARM-VIS	\\
2458494.47998	&	10.92	&	2.84	&	CARM-VIS	\\
2458495.38384	&	-1.63	&	1.50	&	CARM-VIS	\\
2458496.39060	&	2.20	&	1.26	&	CARM-VIS	\\
2458497.37685	&	3.20	&	1.71	&	CARM-VIS	\\
2458510.32361	&	5.17	&	1.23	&	CARM-VIS	\\
2458518.29593	&	0.13	&	4.71	&	CARM-VIS	\\
2458518.31966	&	1.03	&	6.37	&	CARM-VIS	\\
2458518.39467	&	7.48	&	3.70	&	CARM-VIS	\\
2458528.30334	&	-3.74	&	0.97	&	CARM-VIS	\\
2458529.29283	&	5.62	&	1.23	&	CARM-VIS	\\
2458530.31263	&	-7.25	&	1.39	&	CARM-VIS	\\
2458532.31068	&	-8.08	&	1.31	&	CARM-VIS	\\
2458534.30066	&	-4.26	&	1.60	&	CARM-VIS	\\
2458535.30898	&	-2.50	&	1.53	&	CARM-VIS	\\
2458538.29198	&	3.39	&	1.35	&	CARM-VIS	\\
2458539.31096	&	-10.87	&	1.41	&	CARM-VIS	\\
2458540.29356	&	1.70	&	1.48	&	CARM-VIS	\\
2458541.31743	&	-7.09	&	1.79	&	CARM-VIS	\\
2458693.66944	&	3.18	&	1.30	&	CARM-VIS	\\
2458709.66857	&	7.65	&	1.56	&	CARM-VIS	\\
2458710.67818	&	-1.07	&	1.34	&	CARM-VIS	\\
2458714.67355	&	5.16	&	1.46	&	CARM-VIS	\\
2458715.67651	&	-5.36	&	1.33	&	CARM-VIS	\\
2458721.68168	&	3.79	&	2.44	&	CARM-VIS	\\
2458723.68616	&	9.43	&	1.66	&	CARM-VIS	\\
2458725.68713	&	3.79	&	1.41	&	CARM-VIS	\\
2458731.67728	&	-6.95	&	1.33	&	CARM-VIS	\\
2458735.66462	&	-1.37	&	1.81	&	CARM-VIS	\\
2458742.66198	&	-6.07	&	1.79	&	CARM-VIS	\\
2458744.64352	&	1.32	&	1.41	&	CARM-VIS	\\
2458749.67624	&	-4.62	&	2.37	&	CARM-VIS	\\
2458756.59954	&	-9.75	&	1.32	&	CARM-VIS	\\
2458757.64116	&	-0.25	&	1.37	&	CARM-VIS	\\
2458758.61752	&	-0.67	&	1.49	&	CARM-VIS	\\
2458759.62519	&	-1.73	&	1.69	&	CARM-VIS	\\
2458760.62206	&	5.30	&	1.31	&	CARM-VIS	\\
2458761.59804	&	-4.53	&	1.35	&	CARM-VIS	\\
2458762.61542	&	4.19	&	1.14	&	CARM-VIS	\\
2458763.61099	&	-3.39	&	1.58	&	CARM-VIS	\\
2458764.61249	&	3.97	&	1.38	&	CARM-VIS	\\
2458765.55554	&	-2.36	&	1.37	&	CARM-VIS	\\
2458766.55575	&	-2.92	&	1.29	&	CARM-VIS	\\
2458767.59472	&	9.57	&	1.50	&	CARM-VIS	\\
2458769.59518	&	6.21	&	1.80	&	CARM-VIS	\\
2458770.59142	&	-3.15	&	1.77	&	CARM-VIS	\\
2458774.58130	&	3.99	&	1.12	&	CARM-VIS	\\
2458775.57588	&	0.43	&	1.57	&	CARM-VIS	\\
2458777.56053	&	-4.45	&	1.90	&	CARM-VIS	\\
2458781.68039	&	-0.59	&	1.83	&	CARM-VIS	\\
2458783.53425	&	6.72	&	1.16	&	CARM-VIS	\\
2458784.55096	&	-3.91	&	1.43	&	CARM-VIS	\\
2458785.53872	&	7.88	&	1.38	&	CARM-VIS	\\
2458791.65437	&	-10.12	&	3.37	&	CARM-VIS	\\
2458796.53106	&	1.60	&	1.93	&	CARM-VIS	\\
2458801.57402	&	9.55	&	2.00	&	CARM-VIS	\\
2458804.52117	&	0.37	&	1.93	&	CARM-VIS	\\
2457703.50776	&	-5.59	&	4.35	&	CARM-NIR	\\
2457752.42927	&	3.76	&	5.63	&	CARM-NIR	\\
2458001.57240	&	-2.16	&	6.73	&	CARM-NIR	\\
2458043.56937	&	3.72	&	5.84	&	CARM-NIR	\\
2458065.51751	&	-8.03	&	6.32	&	CARM-NIR	\\
2458078.44277	&	2.08	&	4.79	&	CARM-NIR	\\
2458084.46385	&	15.44	&	5.31	&	CARM-NIR	\\
2458109.44945	&	3.26	&	10.63	&	CARM-NIR	\\
2458135.38740	&	10.34	&	5.35	&	CARM-NIR	\\
2458355.59603	&	-3.18	&	6.10	&	CARM-NIR	\\
2458361.63291	&	10.05	&	9.58	&	CARM-NIR	\\
2458379.62915	&	7.16	&	6.05	&	CARM-NIR	\\
2458385.64406	&	-7.43	&	4.43	&	CARM-NIR	\\
2458394.59653	&	-2.50	&	8.63	&	CARM-NIR	\\
2458397.62206	&	-0.87	&	5.17	&	CARM-NIR	\\
2458410.60659	&	-6.14	&	25.51	&	CARM-NIR	\\
2458413.58128	&	-2.61	&	5.46	&	CARM-NIR	\\
2458421.56456	&	0.36	&	16.75	&	CARM-NIR	\\
2458424.67752	&	-0.46	&	9.16	&	CARM-NIR	\\
2458426.49280	&	-2.82	&	7.82	&	CARM-NIR	\\
2458433.54451	&	2.10	&	5.00	&	CARM-NIR	\\
2458434.52562	&	3.76	&	5.71	&	CARM-NIR	\\
2458435.59277	&	-13.85	&	15.38	&	CARM-NIR	\\
2458449.59885	&	-7.41	&	9.93	&	CARM-NIR	\\
2458450.48087	&	1.14	&	5.72	&	CARM-NIR	\\
2458451.46350	&	4.77	&	5.16	&	CARM-NIR	\\
2458454.48317	&	-9.99	&	7.38	&	CARM-NIR	\\
2458470.42305	&	-7.19	&	4.96	&	CARM-NIR	\\
2458474.42722	&	11.65	&	10.78	&	CARM-NIR	\\
2458475.42661	&	-5.16	&	5.13	&	CARM-NIR	\\
2458476.41720	&	12.88	&	3.99	&	CARM-NIR	\\
2458477.41891	&	0.47	&	13.23	&	CARM-NIR	\\
2458478.40329	&	2.85	&	6.48	&	CARM-NIR	\\
2458480.39312	&	2.84	&	5.07	&	CARM-NIR	\\
2458481.38660	&	-5.09	&	10.12	&	CARM-NIR	\\
2458483.40314	&	-1.08	&	8.21	&	CARM-NIR	\\
2458484.37506	&	-3.56	&	5.73	&	CARM-NIR	\\
2458485.40026	&	8.65	&	4.23	&	CARM-NIR	\\
2458486.38572	&	-4.78	&	5.71	&	CARM-NIR	\\
2458487.49941	&	3.96	&	5.86	&	CARM-NIR	\\
2458488.37376	&	-1.29	&	4.74	&	CARM-NIR	\\
2458489.37987	&	-0.66	&	5.23	&	CARM-NIR	\\
2458490.38516	&	9.30	&	6.46	&	CARM-NIR	\\
2458491.41357	&	-9.03	&	4.91	&	CARM-NIR	\\
2458492.37997	&	9.75	&	4.61	&	CARM-NIR	\\
2458492.40359	&	11.84	&	5.42	&	CARM-NIR	\\
2458493.38592	&	-20.43	&	8.08	&	CARM-NIR	\\
2458494.47974	&	-3.03	&	9.71	&	CARM-NIR	\\
2458495.38245	&	2.30	&	5.43	&	CARM-NIR	\\
2458496.35648	&	-39.03	&	24.68	&	CARM-NIR	\\
2458496.39021	&	-0.62	&	5.57	&	CARM-NIR	\\
2458497.37586	&	9.50	&	5.48	&	CARM-NIR	\\
2458510.32343	&	1.48	&	14.66	&	CARM-NIR	\\
2458518.29586	&	-13.49	&	19.23	&	CARM-NIR	\\
2458518.32058	&	-14.25	&	22.85	&	CARM-NIR	\\
2458518.34696	&	-2.86	&	20.58	&	CARM-NIR	\\
2458518.39461	&	-9.08	&	23.16	&	CARM-NIR	\\
2458528.30243	&	-1.30	&	4.42	&	CARM-NIR	\\
2458529.29228	&	8.03	&	4.13	&	CARM-NIR	\\
2458530.31288	&	-5.18	&	10.18	&	CARM-NIR	\\
2458532.31050	&	-5.71	&	5.31	&	CARM-NIR	\\
2458534.30032	&	-10.32	&	5.80	&	CARM-NIR	\\
2458535.30817	&	2.28	&	5.50	&	CARM-NIR	\\
2458538.29222	&	3.20	&	4.42	&	CARM-NIR	\\
2458539.31097	&	-18.41	&	5.25	&	CARM-NIR	\\
2458540.29336	&	-11.13	&	6.17	&	CARM-NIR	\\
2458541.31701	&	-5.54	&	6.73	&	CARM-NIR	\\
2458693.67033	&	1.48	&	4.60	&	CARM-NIR	\\
2458709.66834	&	5.66	&	6.12	&	CARM-NIR	\\
2458710.67867	&	-5.85	&	5.18	&	CARM-NIR	\\
2458714.67398	&	4.15	&	4.84	&	CARM-NIR	\\
2458715.67641	&	-2.47	&	4.53	&	CARM-NIR	\\
2458721.68158	&	29.70	&	15.35	&	CARM-NIR	\\
2458723.68638	&	7.94	&	5.49	&	CARM-NIR	\\
2458725.68696	&	0.28	&	6.22	&	CARM-NIR	\\
2458731.67697	&	-5.42	&	5.47	&	CARM-NIR	\\
2458735.66467	&	-0.67	&	6.70	&	CARM-NIR	\\
2458742.66108	&	-6.05	&	4.92	&	CARM-NIR	\\
2458744.64351	&	-0.73	&	4.47	&	CARM-NIR	\\
2458749.67645	&	-2.32	&	8.68	&	CARM-NIR	\\
2458756.60037	&	0.33	&	5.08	&	CARM-NIR	\\
2458757.64113	&	2.16	&	4.72	&	CARM-NIR	\\
2458758.61806	&	2.00	&	7.48	&	CARM-NIR	\\
2458759.62502	&	-0.68	&	5.89	&	CARM-NIR	\\
2458760.62214	&	2.79	&	4.49	&	CARM-NIR	\\
2458761.59865	&	-10.14	&	4.86	&	CARM-NIR	\\
2458762.61566	&	1.87	&	4.47	&	CARM-NIR	\\
2458763.61115	&	-1.06	&	6.41	&	CARM-NIR	\\
2458764.61238	&	4.06	&	4.52	&	CARM-NIR	\\
2458765.55553	&	-4.36	&	4.45	&	CARM-NIR	\\
2458766.55566	&	-4.20	&	4.43	&	CARM-NIR	\\
2458767.59405	&	7.48	&	4.67	&	CARM-NIR	\\
2458769.59525	&	6.90	&	5.92	&	CARM-NIR	\\
2458770.59144	&	2.49	&	12.43	&	CARM-NIR	\\
2458774.58063	&	9.77	&	3.99	&	CARM-NIR	\\
2458775.57505	&	0.32	&	4.81	&	CARM-NIR	\\
2458777.56011	&	8.96	&	12.66	&	CARM-NIR	\\
2458781.67920	&	4.89	&	14.99	&	CARM-NIR	\\
2458783.53402	&	8.36	&	4.64	&	CARM-NIR	\\
2458784.55132	&	-1.59	&	4.48	&	CARM-NIR	\\
2458785.53831	&	12.58	&	4.39	&	CARM-NIR	\\
2458791.65440	&	-1.44	&	17.75	&	CARM-NIR	\\
2458796.53116	&	-2.64	&	6.39	&	CARM-NIR	\\
2458801.57351	&	5.97	&	7.97	&	CARM-NIR	\\
2458804.52074	&	1.17	&	8.00	&	CARM-NIR	\\
2458747.72635	&	-6.49	&	0.88	&	ESPRESSO	\\
2458747.78703	&	-6.31	&	0.74	&	ESPRESSO	\\
2458747.84737	&	-4.82	&	0.61	&	ESPRESSO	\\
2458747.87837	&	-5.01	&	0.71	&	ESPRESSO	\\
2458747.88888	&	-5.55	&	0.67	&	ESPRESSO	\\
2458749.83137	&	-5.20	&	0.58	&	ESPRESSO	\\
2458749.86178	&	-3.47	&	0.66	&	ESPRESSO	\\
2458749.88207	&	-4.27	&	0.58	&	ESPRESSO	\\
2458750.73535	&	2.05	&	0.90	&	ESPRESSO	\\
2458750.80646	&	3.80	&	0.75	&	ESPRESSO	\\
2458750.83495	&	2.62	&	0.86	&	ESPRESSO	\\
2458750.90479	&	6.75	&	0.81	&	ESPRESSO	\\
2458751.68823	&	6.29	&	1.29	&	ESPRESSO	\\
2458751.86500	&	1.61	&	1.04	&	ESPRESSO	\\
2458753.79784	&	7.56	&	0.89	&	ESPRESSO	\\
2458753.85050	&	7.93	&	0.83	&	ESPRESSO	\\
2458753.87137	&	8.61	&	0.78	&	ESPRESSO	\\
        \noalign{\smallskip}
\hline 
\end{longtable}
\end{center}

\end{appendix}


\end{document}